\documentclass[aps,pre,showpacs,reprint]{revtex4-1}
\usepackage{graphicx}

\usepackage{epsfig}
\usepackage[]{natbib}
\usepackage{xspace}
\usepackage{color}
\usepackage{bbold}
\usepackage{tabu}

\usepackage{amsmath}
\usepackage{amssymb}
\usepackage{mathrsfs}
\usepackage{mathtools}
\usepackage{color}
\usepackage{hyperref}
\usepackage{array}

\makeatletter
\g@addto@macro\bfseries{\boldmath}
\makeatother

\begin{document}
 
  \title{Gas of sub-recoiled laser cooled
atoms described by infinite ergodic theory
}

\author{Eli Barkai}
  \affiliation{Department of Physics, Institute of Nanotechnology and Advanced Materials, Bar-Ilan University, Ramat Gan 52900,
  Israel}
\author{G\"unter Radons}
\affiliation{
Institute of Physics, Chemnitz University of Technology, 09107 Chemnitz, Germany}
\affiliation{
Institute of Mechatronics, 09126 Chemnitz, Germany. }
  \author{Takuma Akimoto}
\affiliation{
Department of Physics, Tokyo University of Science, Noda, Chiba 278-8510, Japan}

  \date{\today}

  \begin{abstract}

The velocity distribution
of a classical gas of  atoms in thermal equilibrium is the normal Maxwell 
distribution. 
It is well known that for sub-recoiled laser cooled atoms L\'evy statistics
and deviations from usual ergodic behaviour come into play.  
Here we show how tools from infinite ergodic theory describe  the cool gas. 
Specifically, we derive the scaling function and the infinite invariant density of a stochastic model 
for the  momentum of laser cooled atoms using two approaches. 
The first is a direct analysis of
the master equation and the second following the analysis 
of  Bertin and Bardou using
the lifetime dynamics. The two methods are shown to be identical, but yield different insights into the problem. 
In the main part of the paper 
we focus  on the  case where the laser trapping is strong, namely the rate of escape
from the velocity trap is $R(v) \propto  |v|^\alpha$ for $v \to 0$
 and $\alpha>1$. 
We  construct a machinery to investigate the time averages of physical observables and their relation
to ensemble averages.  The time averages  are given in terms of functionals of the individual stochastic paths, and here we
use a
generalisation of L\'evy walks to investigate the ergodic properties of the system.
Exploring the energy of the system,
we show that when $\alpha=3$
it exhibits a transition between phases where it is either an
integrable or non integrable  observable,
with respect to the infinite invariant measure. This transition corresponds to very different
properties of the mean energy, and to a discontinuous behaviour of the fluctuations. 
Since previous experimental work showed that both $\alpha=2$ and $\alpha=4$ are attainable
we believe that both phases could be explored also experimentally. 
  \end{abstract}

\maketitle

\section{Introduction}
Laser cooled atoms and molecules  are important for fundamental  and practical applications
\cite{ChuRMP,CTRMP,PhilipsRMP,Shuman}. 
It is well known that L\'evy statistics describes some of  the unusual  properties of  
cooling processes \cite{Bardou,Zoller,CT,Renzoni,GadiRMP}. For sub-recoil laser cooling a special
 atomic
trap in momentum space is engineered.  The most efficient cooling is found when
a mean trapping time, defined more precisely below, diverges \cite{CT}. 
In this sense the dynamics is time-scale-free. The fact that the characteristic time diverges, implies that the processes involved are non-stationary.
Further, in the physics literature they are sometimes  
called non-ergodic.
As is well known ergodicity is a fundamental aspect of statistical mechanics. 

Ergodicity implies that time and ensemble averages of physical observables
coincide. This is found when the measurement time is made long compared to the time scale of the dynamics. However, in the context of sub-recoiled laser cooled 
atoms this time 
diverges, and hence no matter how long one measures, deviations from standard
ergodic theory are prominent. Given that lasers replace heat baths in many cooling experiments, what are the ergodic properties of the laser cooled atoms?
In other words, what  replaces the usual ergodic statistical framework?
While previous work investigated thoroughly the distribution
of momentum \cite{CT},  we highlight the role of the non-normalised quasi-steady
state.
Our goal is to show how tools
of infinite ergodic theory describe the statistical properties
of the ensemble and corresponding  time averages of the laser cooled systems.

 Infinite ergodic theory was investigated by mathematicians \cite{Darling,Aaronson,Zweim}
 and more recently 
in Physics \cite{PRLKorabel,Kessler,Miya,Akimoto2012,Kantz,Burioni,Erez,Sato,PRETakuma,Artuso}. 
 It has a  deep relation
with  weak ergodicity breaking found in the context
of glassy dynamics \cite{WEB,Review}. The terminologies  which might seem 
at first conflicting,  will be discussed briefly
towards the end of the paper.   
Infinite ergodic theory
 deals with a peculiar non-normalised density, describing the long time limit,
 hence 
it  is sometimes called the infinite invariant density. Previous works
 in the field of sub-recoil laser cooling \cite{CT,Bertin}
foresaw this quasi-steady state.
Hence we start with a recapitulation exposing the meaning of the
infinite density using a master equation approach.  We  
show how to use this tool to investigate the ensemble
and   time averages of physical observables and discuss the fluctuations.
In particular we investigate the energy of the system. Since the
atoms are non-interacting, in a classical thermal setting the energy of
the atoms per degree of freedom is $k_B T/2$. 
And this is obtained from the Maxwell velocity distribution,
which is, of course, a perfectly normalised density.
We will show, among other things, that the energy of sub-recoiled 
gas is obtained under certain conditions with a non-normalisable
invariant density.  A transition is exposed in the statistical
properties of the energy when the fluorescence rate $R(v)\propto |v|^{\alpha}$, in
the vicinity of zero velocity, is controlled. Since 
experimental work demonstrates the capability of a variation of
$\alpha$ from
$\alpha=2$ to $\alpha=4$ and at least in principle with
 pulse shaping and Raman cooling
\cite{CT,Reichel,ReichelPHD}
 to other values of $\alpha$, 
the rich phase diagram of ergodic properties we find, seems to us within 
reach of experimental investigation.  

\begin{figure}\begin{center}
\epsfig{figure=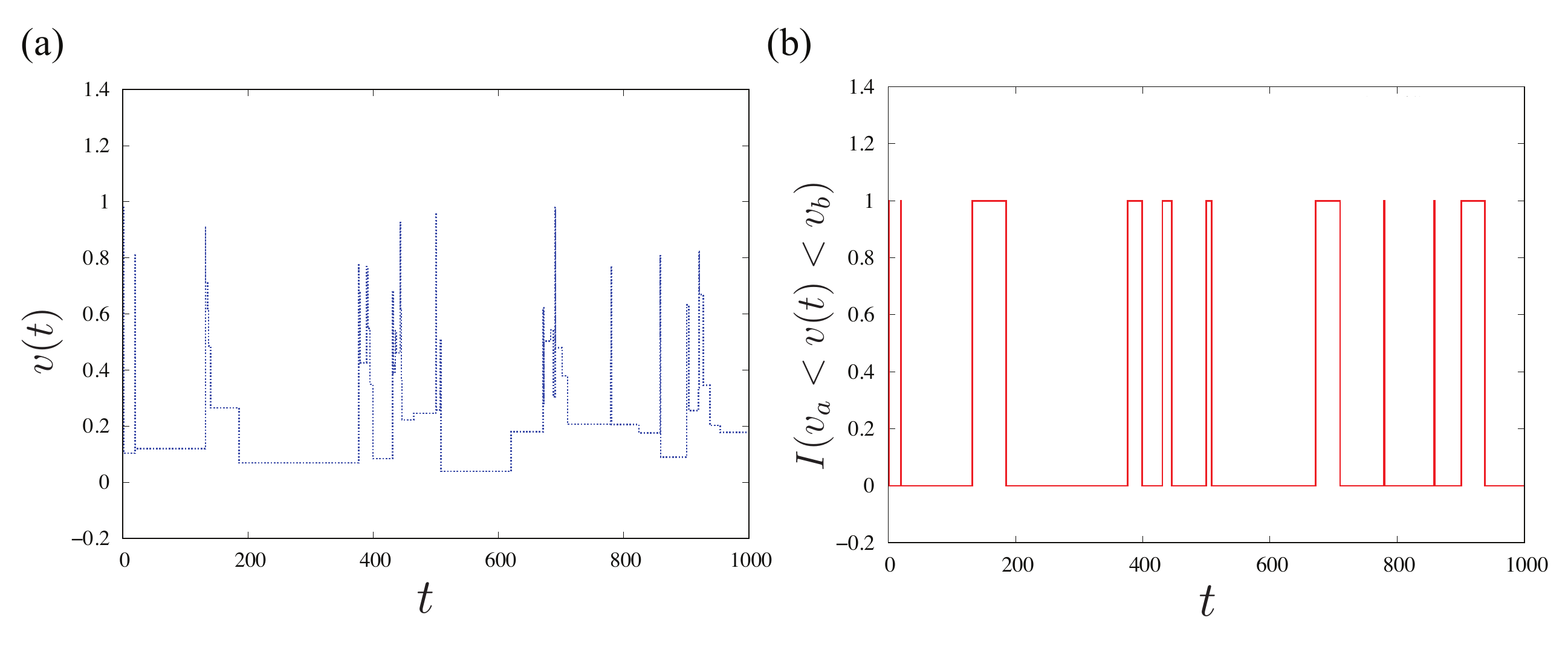, width=0.51\textwidth,trim=0mm 0mm 0mm 0mm, clip}
\end{center}
\caption{
A sample path $v(t)$   exhibits long sticking times whenever the particle is injected to a small velocity. This is due to the trapping mechanism, specifically
 the
vanishing of the collision rate as $R (v) = v^\alpha$ for 
small velocities.  Here we use $\alpha=2 $ and the
parent distribution $f(v)$  is uniform in the interval $(0,1)$.
 We also show the indicator
function which attains the value unity whenever $1/4 < v(t) < 1$, otherwise it is zero. As explained in the text this observable is always integrable
with respect to the infinite density,  a trait crucial for infinite
ergodic theory. 
}
\label{fig1}
\end{figure}

 In this work we are influenced by advances in the statistical theory
of optical experiments, in particular single molecule  tracking.
In this field the removal of the problem of ensemble averaging \cite{MoernerOrrit},
led to new insights into the applications and limitations
 of standard  ergodic theory, for example in the context
of diffusion of single molecules in the cell \cite{Review,Garini}
 and the power-law
distributed sojourn times of dark and bright  states of blinking quantum dots
\cite{Kuno,Stefani}.  
Similarly, here the time averages of a single trajectory
of an atom/molecule in
the process of laser cooling are studied theoretically, as fits this
special issue.
As mentioned, the omnipresent power-law distributed sojourn times, 
are found also
in laser cooled gases, and they describe the times in the momentum trap
in the vicinity of small velocities. These are responsible for the emergence
of the infinite ergodicity framework.  

 A summary of our main results was published in a recent letter
\cite{letter}, while this paper is organised as follows.
We present the model, and analyse the distribution of velocities with a master equation, this gives both a scaling solution and the infinite invariant density,
see Sec. II. Sec. III is devoted to a discussion of both, ensemble and time averages, in generality. We then focus on the energy of the system, developing tools
for analysing the corresponding paths, see Sec IV.  In Sec. V we explore
the Darling-Kac phase, where the observable of choice
is integrable with respect to the non-normalized state corresponding to
 $\alpha<3$.
We then study the non-integrable phase $3<\alpha$, in Sec. VI. We end with open
questions, perspective, and a summary.

\section{Infinite density and scaling solution}

 Let $v>0$ be the speed of the atom under the influence of sub-recoil laser cooling. 
The stochastic process for $v(t)$, presented schematically
in Fig. \ref{fig1},  
is described by the following rules \cite{CT}. 
At time $t=0$ draw the speed $v_1$ from
the probability density function (PDF) $f(v)$.
Momentum is conserved until the atom experiences  a jolt due to the interaction
with the laser field.  
Hence the speed of the atom  will remain fixed for time $\tilde{\tau}_1$.
The PDF of $\tilde{\tau}_1$ conditioned on $v_1$  is  exponential 
\begin{equation}
q(\tilde{\tau}| v ) = \exp[ -\tilde{\tau}/\tau(v) ]/\tau(v).
\label{eqEXP}
\end{equation}
 Here 
$\tau(v)$ is the mean lifetime which depends on the speed of
the particle. 
After time $\tilde{\tau}_1$ we draw a new speed $v_2$  from $f(v)$.   The process is then repeated. Namely we now draw the second waiting time
$\tilde{\tau}_2$ from the PDF in Eq. (\ref{eqEXP}) with an
updated  lifetime  $\tau(v_2)$. This process is then renewed. 
Given $f(v)$ and $\tau(v)$ we are interested in the speed of the particle
at time $t$.  The event of change of velocity is called below a collision
or a jump. 
Here we analyse the velocity distribution using a master equation approach.
A different elegant approach to the problem was considered by Bertin and Bardou 
\cite{Bertin} 
using the dynamics of the lifetimes, and this is presented in Appendix A. 

%

{\em Master equation.}
Let $\rho \left( v,t\right) $ be the PDF of $v$ at time $t$. The repeated
cooling 
process, as described above,
leads to an evolution of $\rho \left( v,t\right) $, which is governed by a
Master equation
\begin{equation}
\frac{\partial \rho \left( v,t\right) }{\partial t}=\int_{0}^{\infty }\left[
W(v^{\prime }\rightarrow v)\rho \left( v^{\prime },t\right) -W(v\rightarrow
v^{\prime })\rho \left( v,t\right) \right] {\rm d}v^{\prime }
\label{Master}
\end{equation}%
containing gain and loss terms, respectively. The independence of
the parent distribution $f(v^{\prime })$ from the previous velocity value $v$
leads to a factorization of the transition rates from $v$ to $v^{\prime }$ 
\begin{equation}
W(v\rightarrow v^{\prime })=R(v)f(v^{\prime }).  \label{transition rates}
\end{equation}%
Using Eq. (\ref{transition rates}) and the normalization $\int_{0}^{\infty
}f(v^{\prime }){\rm d}v^{\prime }=1$ gives 
\begin{equation}
\int_{0}^{\infty }W(v\rightarrow v^{\prime }){\rm d}v^{\prime }=R(v),
\label{jump rate}
\end{equation}
which identifies $R(v)$ as a jump rate, the rate of leaving the state with
velocity $v$, which is the inverse of the lifetime, i.e. $R(v)=1/\tau
(v).$ With these assumptions the Master equation Eq. (\ref{Master})
simplifies to 
\begin{equation}
{\partial {\rho} \over \partial t}  = -{ \rho \over \tau(v)} + f(v) \int_{0} ^\infty { \rho(v',t) \over \tau (v')} {\rm d} v'.
\label{eq01}
\end{equation}
%
%
An invariant density $\rho ^{\ast }\left( v\right)$, which zeros the time derivative on the left hand side of the master equation,  is obtained easily from
Eq.
(\ref{Master})
i.e.
\begin{equation}
\left[ W(v^{\prime }\rightarrow v)\rho ^{\ast }\left( v^{\prime }\right)
-W(v\rightarrow v^{\prime })\rho ^{\ast }\left( v\right) \right] =0
\label{detailed balance}
\end{equation}%
leading to $R(v^{\prime })f(v)\rho ^{\ast }\left( v^{\prime }\right)
=R(v)f(v^{\prime })\rho ^{\ast }\left( v\right) $, \bigskip which is
fulfilled, if e.g. the $v-$dependence on both sides is identical, i.e.,
if $f(v)=R(v)\rho ^{\ast }\left( v\right) C$, \ or 
\begin{equation}
\rho ^{\ast }\left( v\right) =\frac{1}{C}\tau(v) f(v).
\label{invariant density}
\end{equation}
and $C$ is some constant. 
 Below we promote two scenarios for this invariant
solution of Eq. (\ref{eq01}).
The first is well known and it is 
 the case  when $\rho
^{\ast }\left( v\right)$ is normalizable, the second when it is not.
 We will see below that even
if the normalization integral diverges, which happens easily,
then non-normalizable
density $\rho ^{\ast }\left( v\right) $ of Eq. (\ref{invariant density}) is
still meaningful and can be used for calculating certain time
and  ensemble averages.

The  integral in Eq. (\ref{eq01}) represents the process where 
 atom's state $v'$  is shifted due to the laser-atom interaction and the  new velocity $v$
 is drawn from
the PDF $f(v)$.
In \cite{CT}  a uniform model for  
$f(v)$  was investigated,
and hence  (unless stated otherwise)
in our examples below we choose
$f(v) = 1/v_{{\rm max}}$ for $0<v<v_{{\rm max}}$ otherwise it is zero.
In equilibrium, we have
\begin{equation}
\rho^{{\rm eq}} (v) = {\tau (v) f(v) \over Z}
\label{eq02}
\end{equation}
and the normalisation is $Z= \int_0 ^\infty \tau(v) f(v) {\rm d} v$. 
Here the steady state exists in the usual sense and the
normalised density $\rho^{{\rm eq}}(v)$ can be used to predict the 
ensemble and the corresponding time averages of the process. 
Especially Birkhoff's \cite{Birkhoff}  ergodic theory  states that for an observable ${\cal O} [v(t)]$
the time average is equal to the ensemble average
\begin{equation}
\lim_{t \to \infty} {1 \over t}  \int_0 ^t {\cal O} [ v(t')] {\rm d} t' =
\langle {\cal O}(v) \rangle,
\label{eq03}
\end{equation}
where the  ensemble average in equilibrium is
 $$\langle{\cal O} (v) \rangle = \int_0 ^\infty \rho^{{\rm eq}}(v) {\cal O} (v) {\rm d} v. $$ 

However, if 
\begin{equation}
\tau(v) \sim c  v^{-\alpha} \ \ \mbox{ for } 
\  \  v\to 0 
\label{eqseek}
\end{equation}
and $\alpha > 1$ the above standard framework does not work, since $Z$ 
diverges. 
This is precisely the situation for laser cooled atoms where $\alpha=2$ or
$\alpha=4,6$ depending on the specific atom-light interaction
process  \cite{CT,Bertin}. 
Clearly once 
 $v$ becomes small 
then the lifetime  is very long. 
 This is the widely discussed mechanism of sub-recoil cooling, the atoms once slowed down
will have a very long lifetime, and hence remain in the cold state for a long time. The atoms thus pile up
close to zero velocity.
This phase, i.e. $\alpha>1$, is the case  where infinite ergodic theory plays a vital role,  as we will show.

To solve the problem we consider the long time limit and then for $v\neq 0$ we have \cite{CT,Bertin}
\begin{equation}
\rho(v,t) \sim { b \tau(v) f(v) \over t^{ 1 -\tilde{\xi}}} .
\label{eq04}
\end{equation}
This is a quasi-steady state in the sense the numerator is proportional
to the usual $\rho^{{\rm eq}}(v)$
Eq. (\ref{eq02}). 
It is valid for $v$ beyond a small layer around $v\simeq 0$,  which itself shrinks with time to zero, see below.  
It is easy to check that Eq. (\ref{eq04})
is a valid solution  by insertion into Eq. (\ref{eq01}). 
Indeed  the time derivative on the left hand side gives $(\tilde{\xi}-1)\rho(v,t)/t$
which is vanishing when $t\to \infty$,  
  and the right hand side gives zero just like for ordinary steady states.
Here the exponent $0<\tilde{\xi}<1$ is called the infinite density exponent,
and it will  be soon determined 
 and similarly for the constant  $b$. 

 The solution Eq. (\ref{eq04}) 
 is invalid for very small $v$ because it diverges at 
$v\to 0$ in such a way that it is non-integrable since $\alpha>1$.  Following
\cite{CT,Bertin} we seek
a scaling solution
\begin{equation}
\rho(v,t) \sim t^{\tilde{\gamma}} g( t^{\tilde{\gamma}} v) 
\label{eq05}
\end{equation}
which describes the inner region of slow atoms and hence the cooling effect. 
Here $v\propto 1/t^{{\tilde \gamma}}$ and hence the velocity 
is small since $t$ is large. 
Of course the inner and outer  solutions Eqs. (\ref{eq04},\ref{eq05})
must match, and we will exploit this to find the dynamical exponents of this problem $\tilde{\xi}$ and $\tilde{\gamma}$. The latter is what we call the scaling exponent. 
Using Eq. 
(\ref{eq04}) we have $\rho(v,t) \propto v^{-\alpha}$ for $v\to 0$ and this must match the inner solution hence we have
\begin{equation}
 g(x) \propto x^{-\alpha}  \  \ \mbox{for} \ x>> 1. 
\label{eq05a}
\end{equation}
Note that $g(x)$ is integrable namely it can be normalised when $\alpha>1$.

\begin{widetext}
We insert Eq.
(\ref{eq05})
in Eq. (\ref{eq01}) 
to find an equation for $g(x)$. Some thought is required with respect to the
upper limit of integration that stretches to infinite velocities on the right hand side of Eq. (\ref{eq01}). However,
the scaling function does not describe large velocities, in fact velocities of order $v\propto v_{{\rm max}}$ 
are modelled by the quasi-steady state, Eq. (\ref{eq04}),
 as mentioned.
 Let us, however, pretend that we do not know this  and see how it comes out
of the master equation. We recall that $t$ is large, and
 replace the upper limit of 
integration over velocity by an upper cutoff $h t^\beta$. We call $\beta$ the cutoff exponent.
 We then find
using Eqs. (\ref{eq01}) and (\ref{eq05}) 
\begin{equation} 
\tilde{\gamma} t^{\tilde{\gamma}-1}\left[ g(t^{\tilde{\gamma}} v) + 
t^{\tilde{\gamma}} v g' ( t^{\tilde{\gamma}} v) \right]
\simeq
- {t^{\tilde{\gamma}} g(  t^{\tilde{\gamma}} v) \over \tau(v) }
+ f(v) \int_0 ^{ h  t^\beta }  { t^{\tilde{\gamma}} g( t^{\tilde{\gamma}}v)
     \over \tau(v) }{\rm d} v,
\label{eq06}
\end{equation}
where $g'(x)$ is the derivative of $g$ with respect to $x$. 
Clearly the  natural scaled variable is  $x= t^{\tilde{\gamma}} v$.
 We now  realise that only the small $v$ behaviour of $\tau(v)$ determines the properties of the scaling solution and hence $g(x)$.
 This as already pointed out 
is because the particles pile up close to zero velocity and
hence only the small $v$ limit matters.
In contrast note that we need the full structure of $\tau(v)$ to describe
the outer solution Eq. 
(\ref{eq04}). 
 Hence now we replace  $\tau(v) 
\rightarrow c v^{-\alpha}$  
in Eq.
(\ref{eq06})
and after change of variables we find
\begin{equation}
\tilde{\gamma} t^{\tilde{\gamma} - 1} \left[ g(x) + x g'(x)\right] =
- t^{\tilde{\gamma} - \tilde{\gamma} \alpha} {1 \over c}  x^\alpha g(x) + t^{-\tilde{\gamma} \alpha} f\left( { x \over t^{\tilde{\gamma}}} \right) \int_0 ^{h t^{\beta +\tilde{\gamma}}} g(x) x^\alpha {\rm d} x .
\label{eq07}
\end{equation}
Recall the large $x$ behaviour of $g(x)$  
Eq. (\ref{eq05a}) which gives $g(x) x^\alpha\to \mbox{const}$ and hence we may 
find the long time limit of the integral  in Eq. (\ref{eq07})  and get
\begin{equation}
\tilde{\gamma} t^{\tilde{\gamma} - 1 } \left[ g(x) + x g'(x)\right]
= - t^{\tilde{\gamma} - \tilde{\gamma}a~\alpha} ~ {x^\alpha g(x) \over c}  + {\cal N} t^{ \beta+ \tilde{\gamma} - \tilde{\gamma}~\alpha}. 
\label{eq09}
\end{equation}
Here we used $\lim_{t \to \infty} f(x/t^{\tilde{\gamma}})= f(0)$ 
which is a constant not equal zero,  further  $f(0)$ is swallowed in ${\cal N}$ with
other constants and hence is not presented explicitly in Eq. (\ref{eq09}). 
Eq. (\ref{eq09})
 should be time independent hence we can find  the exponents of the 
problem, $\tilde{\gamma} \alpha=1$ and $\beta=0$. The latter is expected since
the cutoff of velocity is of order $t^0$, i.e. $v_{{\rm max}}$. 
The constant ${\cal N}$ is eventually determined from normalisation, see below.

\begin{figure}\begin{center}
\epsfig{figure=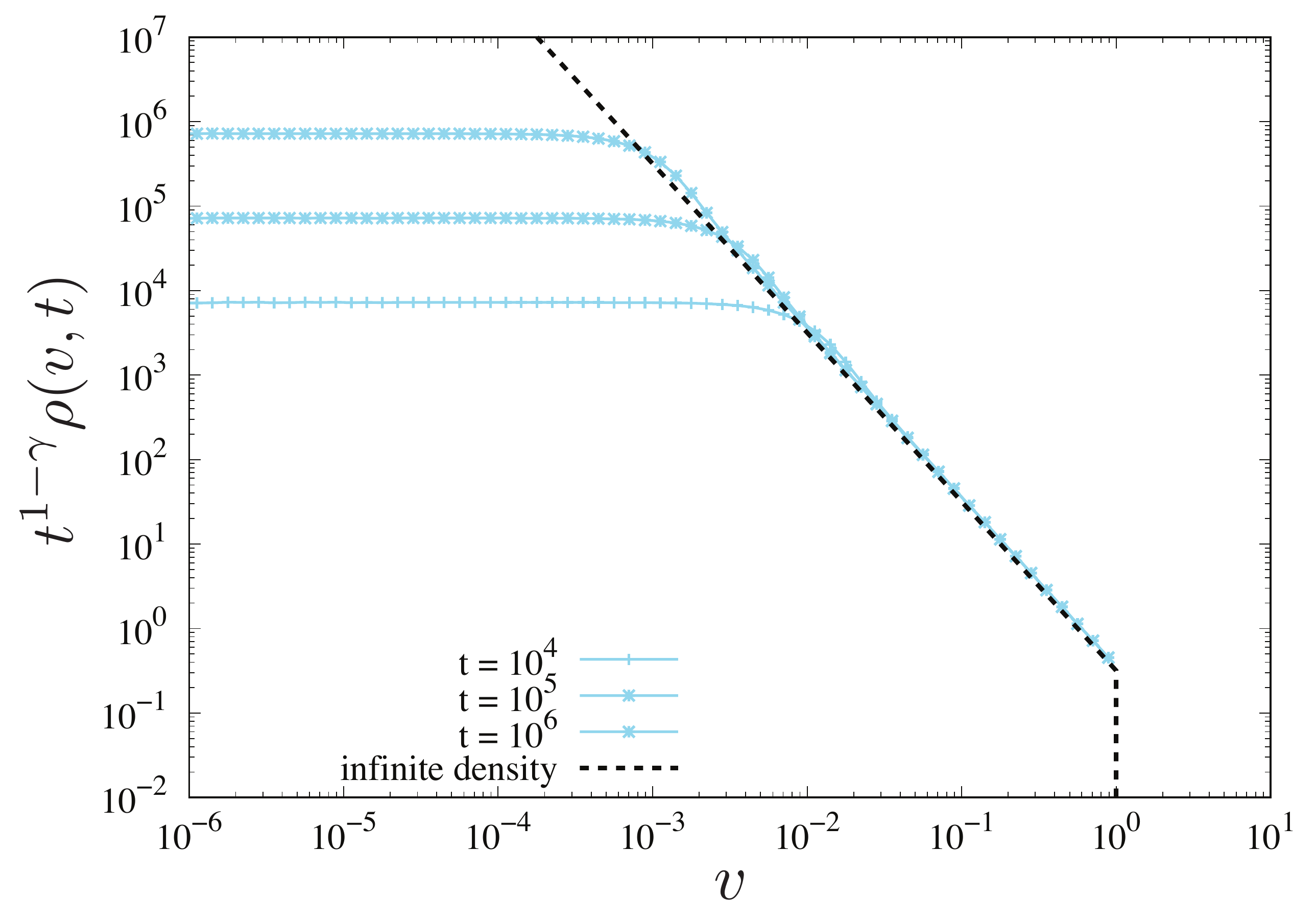, totalheight=0.30\textheight, width=0.55\textwidth,trim=0mm 0mm 0mm 0mm, clip}
\end{center}
\caption{
The scaled PDF of velocity $t^{1-\gamma} \rho(v,t)$ versus $v$ for increasing times as indicated in the figure. A cooling process with a uniform
 parent velocity PDF $f(v)$ the maximum allowed velocity being unity is 
considered. The  collision rate is $R(v) =v^2$ so  $\alpha=2$ and hence $\gamma=1/2$. The data in the long time limit converges to the infinite density Eq.
(\ref{eq12})
presented with a  dashed black line.
}
\label{fig2}
\end{figure}

\begin{figure}\begin{center}
\epsfig{figure=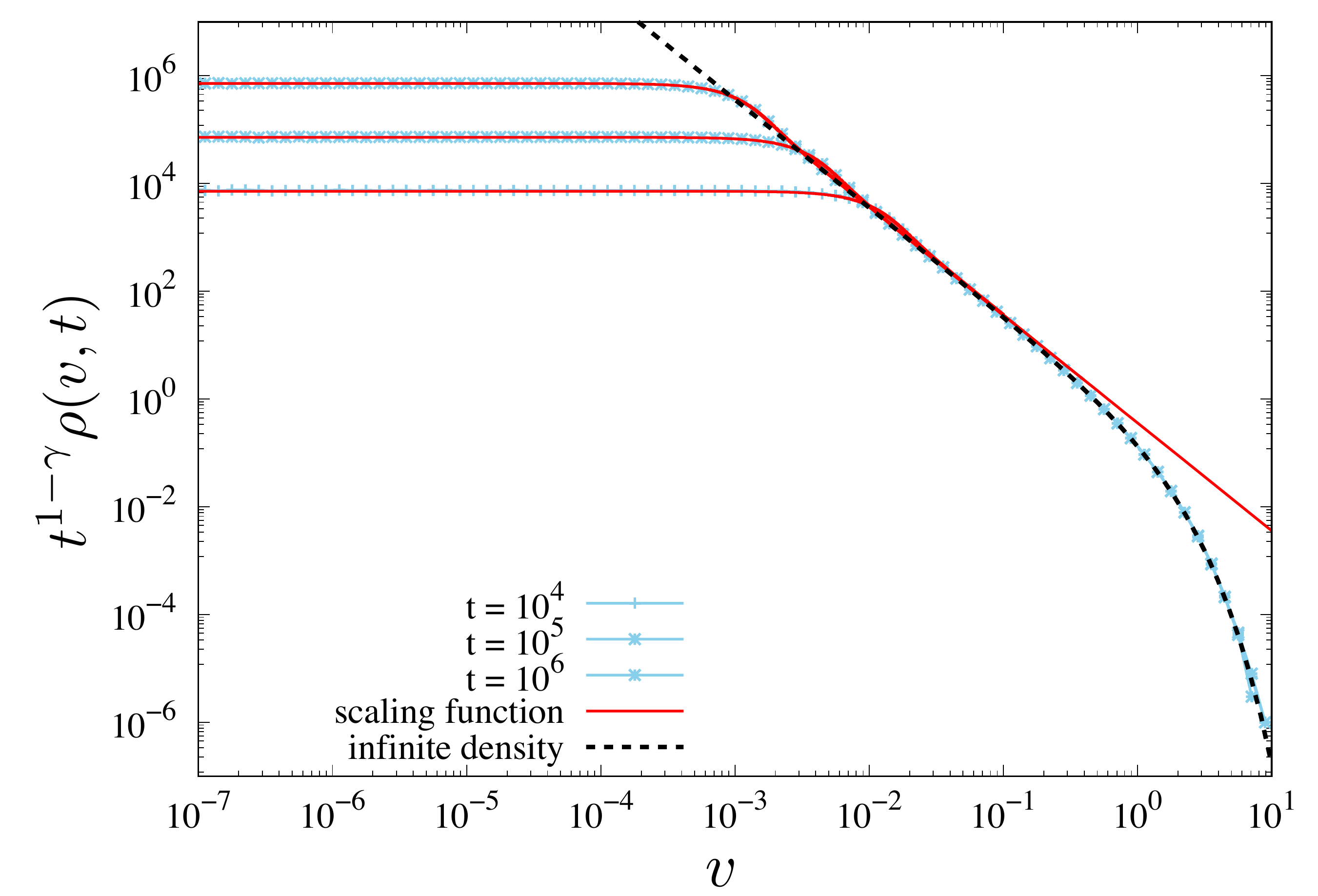, totalheight=0.30\textheight, width=0.55\textwidth,trim=0mm 0mm 0mm 0mm, clip}
\end{center}
\caption{
Similar to Fig. \ref{fig2} we show the scaled PDF of velocity $t^{1-\gamma} \rho(v,t)$ versus $v$, however, here we simulated  a process with an exponential parent velocity PDF $f(v)$ with mean unity, the   collision rate is unchanged
and hence $\alpha=2$. 
The data in the long time limit converges to the infinite density 
Eq. (\ref{eq12})
shown as a  dashed black line. For finite times, we see expected
 deviations between the infinite invariant density and the numerical solution,
which are obvious and clearly visible for small velocities. 
For not too large velocities, i.e. velocities much smaller than the mean of $f(v)$, which is $1/2$ in this example,  
 the scaling solution, 
Eq. (\ref{eq05}) with
Eq. (\ref{eq11})
(red line), nicely matches the data. We see that both,
the scaling solution and the infinite density, describe the distribution of velocities, for $t=10^4$. For intermediate velocities the two solutions match. 
Comparing with  Fig. \ref{fig2} we see
that the sharp cutoff of
the infinite density  (due to the uniform PDF of velocities) is replaced
here 
with an exponential decay.}
\label{fig3}
\end{figure}

 We now use the scaling exponent 
\begin{equation}
\gamma \equiv {1\over \alpha}
\label{EqGamAlp}
\end{equation}
and 
from now  on $\tilde{\gamma} = \gamma$. This exponent is in the range $0<\gamma<1$, and it describes the fat tail  PDF of the waiting time within the momentum
 trap, 
$\phi_1(\tilde{\tau}) \propto\tilde{\tau}^{ - 1 + \gamma}$ (see details below, this PDF is found via averaging the exponential PDF $q(\tilde{\tau}|v)$
from Eq. (\ref{eqEXP})  over the random velocities). Here we see
 that the mean waiting time diverges,
and hence as mentioned in the introduction the process is scale free. 
Generally, infinite ergodic theory is related to such processes, where the mean
of the microscopic time scale diverges. 

 The scaling function is thus determined by
the following  equation
\begin{equation}
\gamma \left[ g(x) + x g'(x) \right] = - { x^{1/\gamma} g(x) \over c}  + {\cal N}.
\label{eq10}
\end{equation}
The substitution  $y(x)=x g(x)$ helps simplifying this equation
even further, and eventually  the solution reads \cite{CT}
\begin{equation}
g(x) = { {\cal N} \over \gamma x} \exp\left( - {x^{1/\gamma} \over c}  \right)\int_0 ^x \exp\left( {z^{1/\gamma} \over c}  \right) {\rm d} z. 
\label{eq11}
\end{equation} 
%
With the help of the generalized Dawson integral $F(p,x)=\exp
(-x^{p})\int\limits_{0}^{x}\exp (y^{p})\rm{d}y$, which according to
\cite{Dawson}  can be expressed by the confluent hypergeometric function $%
M(a,b,z)$, the scaling function can be written as $g(x)=\frac{N}{\gamma }%
\frac{c^{\gamma }}{x}F(\frac{1}{\gamma },\frac{x}{c^{\gamma }})=\frac{N}{%
\gamma }M(1,1+\gamma ,-\frac{x^{1/\gamma }}{c})$, which 
agrees with the result in \cite{CT}. 
${\cal N}$ is determined from normalization
$\int_0 ^\infty g(x) {\rm d} x=1$
and we find 
from formula $7.612(1)$  from 
\cite{GRAD}
\begin{equation}
\frac{{\cal N}}{\gamma} =\frac{1}{\Gamma (1+\gamma )}\frac{\sin (\pi \gamma )}{\pi \gamma } {1 \over c^\gamma} .
\label{eq11a}
\end{equation}

As announced,
the infinite density exponent $\tilde{\xi}$   can be found
by matching the inner and outer solution.
Specifically 
the left part of the outer solution
Eq. (\ref{eq04}) is  $\rho(v,t) \propto v^{-\alpha} / t^{1 - \tilde{\xi}}$
is matched with  the right part of the inner solution Eqs. 
(\ref{eq05},\ref{eq05a}) give $\rho(v,t) \propto v^{-\alpha}/t^{1 - \gamma}$.
resulting in
\begin{equation}
\tilde{\xi}=\gamma. 
\end{equation}
Given ${\cal N}$ in Eq. (\ref{eq11a})
  we can determine $b$ in Eq. (\ref{eq04}). Using integration by parts Eq.
(\ref{eq11}) 
\cite{Dawson}
 yields 
$g(x) \sim {\cal N} c /x^{1/\gamma}$ for
$x \to \infty$.
 So from the scaling solution 
Eq. (\ref{eq05})
we have for large $x=v t^\gamma$, 
\begin{equation}
\rho(v,t) \sim { {\cal N} c \over t^{1-\gamma} v^{1/\gamma}}.
\end{equation} 
This is matched to 
the small $v$ behaviour of the infinite density 
solution 
Eq. (\ref{eq04})
which gives using Eq. 
(\ref{eqseek})
\begin{equation}
\rho(v,t) \sim { b  f(0) c \over t^{1-\gamma} } {1 \over v^{1/\gamma}}. 
\end{equation} 
Thus we find the constant $b$ given by 
$b f(0)= {\cal  N}$.  

To summarise we find that 
\begin{equation}
\boxed{
\lim_{t \to \infty} t^{1-\gamma} \rho(v,t) = {\cal N}  \tau(v) {f(v) \over f(0)}  ={\cal I}_v(v). 
}
\label{eq12}
\end{equation}
The function ${\cal I}_v(v)$ is non-normalisable since 
${\cal I}_v(v) \propto v^{-1/\gamma}$ for $v\to 0$  and $1/\gamma=\alpha>1$. This is why it is
called an infinite density. The non-normalizability is hardly surprising since
we take a perfectly normalised PDF $\rho(v,t)$ 
 and multiply it by $t^{1-\gamma}$ hence
the area under the product  will diverge in the long time limit. Still the
function ${\cal I}_v(v)$ can be used to calculate ensemble and time
averages, as we will show below. 

Figs. \ref{fig2},\ref{fig3}  demonstrate the main
results of this section using finite time simulations of the process.  
In Fig. \ref{fig2} 
simulations are presented for the uniform  velocity PDF $f(v)$
 and we use $v_{\rm max}  = 1$ while for the rate function we use  $c=1$.
Unless stated otherwize  these parameters will be used in all the figures of this paper. 
In Fig. \ref{fig3}
 we show the case where the parent velocity 
distribution is exponential. While the
scaling  solution is not sensitive to the shape of $f(v)$,
 the infinite density is, as demonstrated in the figures.

The distribution for  very small velocity is described by
Eq. (\ref{eq05}) with the scaling function
$g(x)$, Eq. (\ref{eq11}). This scaling solution as a stand alone 
is not a sufficient description of a cooled system for the following reason.
From Eqs. (\ref{eq05},\ref{eq05a})  we have $\rho(v,t) \propto
 v^{-\alpha}$ for large $v$  and
for now
 let us focus on the realistic choice  $\alpha=2$.
We find the awkward situation that the second moment of $v$, namely the mean
kinetic energy would be infinite, due to the fat-tail of the solution, 
 which is not what we expect from
a cold system. Thus the scaling solution $g(x)$  describes the pile up
of particles close to zero velocity, but it also exhibits an unexpected
 heavy power law tail  for large velocities. 
The infinite
density describing the outer region cures this problem, in the sense
that it describes the large velocity 
cutoff, see Figs. \ref{fig2},\ref{fig3}. From here we see that for calculating, for instance the mean kinetic
energy, we need the infinite density ${\cal I}_v (v)$. The technical  details
of this calculation will be presented below. 
 Further, while the scaling function
cures the non-normalisable trait of the infinite density which stems from
its small $v$ behaviour,  we find
that also the complement is true: 
the infinite density cures the unphysical divergence of the kinetic  energy, 
found using the scaling solution, which is due to its unphysical large
 $v$ properties.
 In short  we need
both tools to describe the velocity distribution. 

 Finally, there is experimental evidence for the scaling solution.
According to Eq. 
(\ref{eq05})
the width of the velocity distribution, is proportional to $t^{-\gamma}$ and
hence to $t^{-1/\alpha}$ \cite{Reichel,CT1}.
 Experimentally one may  control the shape of the rate function $R(v)=1/\tau(v)$
in the vicinity of zero, and engineered experiments provided the values $\alpha=2$ and $\alpha=4$. Note that $\alpha=2$ corresponds to an analytical  expansion of $R(v) \propto v^2$ around
its minimum at $v=0$. Hence the theory Eq. (\ref{eq05})
 predicts that the full width at half maximum of the velocity PDF decays like
$t^{-1/2}$ for $\alpha=2$  and as $t^{-1/4}$ for $\alpha=4$. 
Both these behaviours were indeed observed experimentally,  up to times
of order $20$ milliseconds, which given the lifetime of the atom, is considered
very long in this field.  This cooling experiment from $1995$, 
 using  Cesium, reports below 
$3$ nano-Kelvin temperatures and due to the non-Gaussian  nature of the momentum packet, 
temperature is  defined by the width of 
the momentum
 distribution and not by its variance.  
This reality is also consistent with quantum Monte Carlo simulations \cite{CT}. 
In a later experiment, 
the momentum distribution of Helium 
was recorded in full agreement with the statistical method
\cite{CT,Saubamea}. 
As far as we know, so far,  the infinite density was 
not explored  experimentally. 

\section{Ensemble and time averages}

 We now consider the long time limit of averaged observables,
the latter  denoted ${\cal O}[v(t)]$,
so the observable is a function
of the stochastic process $v(t)$. In usual statistical physics
 both,  the ensemble average and the time average of
such observables, are investigated 
 and we do the same here. Examples are
the indicator function
${\cal O}[v(t)] = I[v_a<v(t)<v_b]$ and the
kinetic energy of the particle ${\cal O}[v(t)] = v^2(t) = E_k(t)$ and 
$m/2=1$ with $m$ the mass of the atom.
   The indicator function equals unity, if the condition $v_a<v(t)<v_b$ is true, otherwise it is zero, namely it is an observable that switches at random times between $1$ and $0$, thus this  observable represents a dichotomous
process, see Fig. \ref{fig1}. 
The kinetic energy $E_k(t)$  is an observable that needs no 
introduction. 
 The ensemble average is
\begin{equation}
\langle {\cal O}(t) \rangle = \int_0 ^\infty  {\cal O}(v) \rho(v,t){\rm d} v 
\label{eqave01}
\end{equation}
and hence we find using 
Eq. 
(\ref{eq12})
\begin{equation}
\lim_{t \to \infty} t^{1 - \gamma} \langle {\cal O}(v) \rangle = \int_0 ^\infty
{\cal O}(t) {\cal I}_v(v) {\rm d} v. 
\label{eqave02}
\end{equation}
In this sense the infinite invariant  density ${\cal I}_v(v)$
 replaces the standard invariant density $\rho^{{\rm eq}}(v)$. The formula is valid provided that the integral
is finite and such an observable ${\cal O}(v)$ is called integrable with respect to the infinite density. 
Examples are the indicator  function and the kinetic energy
\begin{equation}
\lim_{t \to \infty} t^{1-\gamma} \langle I(v_a<v(t)<v_b) \rangle =
\int_{v_a} ^{v_b} {\cal I}_v(v) {\rm d} v,
\ \ \
\mbox{and} \ \ \ \
\lim_{t \to \infty} t^{1 - \gamma} \langle E_k(t) \rangle = \int_0 ^\infty v^2 {\cal I}_v (v) {\rm d} v \ \ \mbox{for} \ \ \alpha<3. 
\label{eqave03}
\end{equation}
The former shows that the number of particles in the interval $v_a<v<v_b$ is 
shrinking in time, provided that $0<v_a$ and the latter indicates
that the energy
of the system is decaying similarly,  like $ \langle I(v_a<v(t)<v_b) \rangle \propto \langle E_k(t) \rangle \propto 1/t^{1-\gamma}$. Since ${\cal I}_v (v) \propto
v^{-\alpha}$ for $v \to 0$ we see that the second
integral exists only when $\alpha<3$, hence the kinetic energy is an integrable observable only
if $\alpha<3$. If this condition does not hold,
the kinetic energy is called a 
non-integrable observable and then other rules apply, see below. 
Roughly speaking integrable observables cure the non-normalisable
trait of the infinite density found for $v\to 0$,
 and importantly these integrable
observables are very basic to physical systems, for example the energy. 

 We now consider the time average of an integrable observable
\begin{equation}
\overline{{\cal O}}(t) = {1 \over t}  \int_0 ^t {\cal O} [v(t')] {\rm d} t'.
\label{eqave04}
\end{equation}
In standard ergodic theory $\lim_{t \to 
\infty} \overline{{\cal O}}(t) = \langle {\cal O} \rangle_{{\rm eq}}$.
In the current case the dimensionless variable 
 $\Upsilon=\gamma \overline{O}(t)/\langle {\cal O}(t) \rangle$ remains 
random in the long time limit and as shown below it satisfies the 
Darling-Kac theorem \cite{Darling} provided the observable is integrable. 
Here the mean of $\Upsilon$ in the long measurement time limit is unity  
and importantly in that limit its PDF is time independent. 
First let us consider
the ensemble mean, namely we consider an ensemble of paths and average
over time and then over the ensemble
\begin{equation}
\langle {\overline {\cal O}}(t) \rangle = \left\langle {1 \over t}
 \int_0 ^t {\cal O}[v(t')] {\rm d} t'  \right\rangle = 
{1 \over t}
 \int_0 ^t \int_0 ^\infty {\cal O}(v) \rho(v,t') {\rm d} v {\rm d} t'.
\label{eqave05}
\end{equation}
Here we switched the order of the time and ensemble averages. Considering the
long time limit and using Eq. 
(\ref{eq12})
\begin{equation}
\boxed{
\langle {\overline {\cal O}} (t) \rangle \sim
 {1  \over t} \int_0 ^t  {\rm d} t' { \int_0 ^\infty {\cal O}(v) {\cal I}_v (v) {\rm d} v \over t'^{1-\gamma}} = { \int_0 ^\infty {\cal O}(v) {\cal I}_v (v) {\rm d} v \over \gamma t^{1-\gamma}} 
}
\label{eqave06}
\end{equation}
where we used $0<\gamma<1$. Hence we conclude that
\begin{equation}
\boxed{
\lim_{t \to \infty} { \langle \overline {\cal O}(t) \rangle  \over  \langle {\cal O}(t) \rangle} ={ 1 \over \gamma} .
}
\label{eqave07}
\end{equation}
Thus we established a relation between the time average and the 
ensemble average. The latter is obtained using the infinite density ${\cal I}_v(v)$
 and thus this invariant
density is not merely a tool for the calculation of the ensemble average
but rather it gives also information on the time average. More precisely 
as discussed below $\Upsilon$
 exhibits universal statistics independent of the observable,
provided that it is an integrable observable.  The $1/\gamma$ factor in Eq. (\ref{eqave07}) stems from the time averaging and for example if $\alpha=1/\gamma=2$ we have
$\langle \overline{{\cal O}}(t)\rangle \sim  2  \langle {\cal O}(t) \rangle$,
or since both sides of this equation actually go to zero,
 a more refined 
statment is  $\lim_{t \to \infty} \langle \overline{{\cal O}}(t) \rangle / \langle {\cal O}(t) \rangle= 2$.

\section{Kinetic energy}

We consider the time average of the kinetic energy of the atom
\begin{equation}
\overline{E}_k(t) = \overline{ v^2 }(t) = {1 \over t} \int_0 ^t v^2 (t') {\rm d}t'
\label{ctrw00}
\end{equation}
in the limit of long times. 
While we focus on a particular observable, the theory presented below is actually rather general. 
As mentioned, depending on the value of $\alpha$,
the observable $v^2(t')$
can be either integrable with respect to the infinite
density, or not. Hence, here we will develop a general theory, for the time
averages,  valid in principle  wether the observable is integrable or not.
To do so, we
use tools from random walk theory \cite{MetzKlaf,Kutner},
in particular certain types
of L\'evy walks \cite{Denisov,Shinkai,Albers}.

 We focus on the numerator of Eq. (\ref{ctrw00})
which is a functional of the velocity path,
\begin{equation} 
{\cal S}(t) = \int_0 ^t v^2 (t') {\rm d} t'.
\label{ctrw01}
\end{equation}
Since we have no potential energy, ${\cal S}(t)$ is the action, which is increasing with time and $\overline{E}_k (t)={\cal S}(t) /t$.  Since the speed of particle is a constant between collision
events  
\begin{equation}
{\cal S}(t) = \sum_{i=1} ^{N(t)} (v_i)^2 \tilde{\tau}_i +  (v_{N(t)+1})^2 t_B(t). 
\label{ctrw02}
\end{equation}
Here $v_1$ is the velocity drawn at the start of the process, $v_2$ the
velocity drawn after the first collision, etc. The times $\tilde{\tau}_i$
are the times between collision events. $t_B(t)$ is the time elapsing between
measurement time $t$ and the last event in the sequence, also called
the backward recurrence time \cite{Godreche2001,Wanli}.
 Here
we have the constraint 
\begin{equation}
\sum_{i=1} ^{N(t)}  \tilde{\tau}_i + t_B(t) = t,
\label{eqCONS}
\end{equation}
see schematics in Fig.
\ref{fig4a}.
 Finally, $N(t)$ is the random number of collisions
until time $t$. 

 Recall, that at each collision event we draw $v_i$ from the parent PDF
$f(v)$ and then the waiting time $\tilde{\tau}_i$ from 
Eq. (\ref{eqEXP})
$q(\tilde{\tau}|v_i) = R(v_i) \exp[ - \tilde{\tau} R(v_i)]$,
which is the conditional PDF of the waiting time, given a velocity $v_i$.
 Hence the waiting
times are not identically distributed, unless the rate is a constant. 
Further in this section we will assume that
\begin{equation}
 f(v)= {1 \over v_{{\rm max}}}  \ \ \mbox{when} \ \   0<v<v_{{\rm max}}
\label{eqFV}
\end{equation}
and zero otherwise. 
Thus we consider a uniform distribution of the speed, and further
the rate is
\begin{equation}
 R(v) = {v^\alpha \over c} 
\label{eqEVC}
\end{equation}
which is the inverse of the mean decay time
 $\tau(v)$ in Eq. 
(\ref{eqseek}).
 Here  $c$ is a constant with units
of time times speed to the power of $\alpha$.  

 We may write $s_i=(v_i)^2\tilde{\tau}_i $ and similarly $s_B(t)= (v_{N(t)+1})^2 t_B(t)$ 
and then 
\begin{equation}
{\cal S}(t) = \sum_{i=1} ^{N(t)} s_i +  s_B(t).
\label{ctrw03}
\end{equation}
At first this appears to be a problem of the summation of $N$ random variables,
which is classical in many fields, in particular in the theory of random walks.
However, here $N(t)$ is random, and is determined by the sequence of waiting times, which in turn are  correlated with the jump size, i.e. the increments of 
$s_i$ of
of ${\cal S}(t)$ Eq. (\ref{ctrw03})  
and also the $\tau_i$s in Eq. (\ref{eqCONS}) are
 constrained by the measurement time
$t$. 
 In particular
when $\alpha>1$ the statistics is very different from normal.

{\bf Remark:} In what sense are we dealing here with a generalized L\'evy walk?
 The simplest form of the L\'evy walk deals with the
displacement of a particle $-\infty < X(t)<\infty$ which is 
given by $X(t) =\sum_i ^{N(t)} v_i \tilde{\tau_i} + v_{N+1} t_B(t)$. Here
 the velocity  $v_i$ is say either
$+1$ or $-1$ with probability half (unbiased random walk) while the travel time PDF is fat tailed.  In our study
$S(t)>0$ plays a role similar to $X(t)$ of a biased  L\'evy walk.
The main difference is the following. For standard L\'evy walks
the distribution of the flight times  $\tilde{\tau}$ is independent
from the value of the velocity, i.e. the joint PDF of velocity
and travel times is a product of the marginal PDFs.
 Here, we have a different situation, as the velocity of the atom after the jump event is correlated with the random time of flight by a law determined with
$R(v)=1/\tau(v)$. 
However,  as stated in the abstract,
 the action $S(t)$ is performing a generalised type
of  L\'evy walk, which is  analysed below. 

\subsection{PDF of action increments and
waiting times}

 The joint PDF of $s,\tilde{\tau}$  and $v$
is
\begin{equation}
\phi_3 (s,\tilde{\tau},v) =  \delta (s - v^2 \tilde{\tau})q (\tilde{\tau}| v)
f(v)
\label{ctrw04}
\end{equation}
where $f(v)$ and $q(\tilde{\tau}|v)$ are defined in Eqs. 
(\ref{eqEXP},\ref{eqFV}) respectively. 
Here the delta function comes from the definition of one step,
i.e. given a specific velocity and waiting time the increment of the action 
is fixed.  The subscript $3$ indicates that we are dealing here with three
random variables. We already mentioned that the action  increments 
and the waiting times are correlated and hence to solve the problem
we need the joint PDF of $s$ and $\tilde{\tau}$. This is obtained from
integration of Eq. (\ref{ctrw04}) over $v$ which gives
\begin{equation}
\phi_2 (s, \tilde{\tau}) = { 1 \over 2 v_{{\rm max}} \sqrt{ s \tilde{\tau} }} R \left(  \sqrt{ { s \over \tilde{\tau} }} \right) 
\exp\left[ - \tilde{\tau}  R \left(  \sqrt{ { s \over \tilde{\tau} }} \right) \right]
\label{ctrw05} 
\end{equation} 
when $0\le s \le  v_{{\rm max}} ^2\tilde{\tau}$ and if this condition is not valid,
the joint PDF is equal zero. 
In the context of random walk theory such joint distributions are
investigated in the context of {\em  coupled}  continuous time random walks
\cite{Denisov,Albers,KBS,Miya1PRE,Miya2JSP,Aghion}
since the increment $s$ and waiting times $\tilde{\tau}$ are correlated,

Further integrating over $s$ we get the marginal PDF of the waiting times.
\begin{equation}
\phi_1 (\tilde{\tau}) = { c^{1/\alpha} \over \alpha v_{{\rm max}}} \tilde{\tau}^{-1 - 1/\alpha}  \gamma\left( 1 + {1 \over \alpha} , { v_{{\rm max}} ^\alpha \tilde{\tau} \over c} \right), 
\label{ctrw06} 
\end{equation}
where $\gamma(.,.)$ is the lower incomplete Gamma function.
For large $\tilde{\tau}$ we get a power law decay
\begin{equation}
\phi_1 (\tilde{\tau} )  \sim  \mbox{const} \  \tilde{\tau}^{ - 1 -1/\alpha} 
\label{ctrwaa}
\end{equation}
hence if
$\alpha>1$ the mean waiting time diverges and  $\mbox{const} = c^{1/\alpha} \Gamma(1 + 1/\alpha) / ( \alpha v_{{\rm max}})$. As is well known \cite{WEB,Review}  the divergence of the mean waiting time signals special ergodic properties, since no matter how long we measure
we cannot exceed the mean time, and hence ergodicity in its usual sense
is broken.  At short waiting times we have
$\phi_1(\tilde{\tau})\sim v_{{\rm max}}^\alpha / [(1 + \alpha) c]$. 

The marginal PDF of the action increment is found integrating Eq.
(\ref{ctrw04}) over $\tilde{\tau}$ and $v$. 
We use the notation that the argument in the PDF defines the 
variable under study and in this case
it is easy to show that
\begin{equation}
\phi_1 (s) = { 1 \over v_{{\rm max}}} \int_0 ^{v_{{\rm max}}} { v^{\alpha-2} \over c} \exp\left( - { v^{\alpha-2} s  \over c} \right) {\rm d} v. 
\label{ctrw07} 
\end{equation}
Then for the example $\alpha=2$ we find $\phi_1(s) = \exp(-s/c)/c$, namely an exponential decay.
More generally the mean of the action increments
$\langle s \rangle = \int_0^{\infty} s \phi_1(s) {\rm d} s$
 is an important
quantifier 
and Eq.  (\ref{ctrw07}) gives
\begin{equation}
\langle s \rangle = 
\left\{ 
\begin{array}{c c}
{ c v_{{\rm max}} ^{2-\alpha} \over 3 -\alpha} & \alpha<3 \\
\infty & \alpha > 3
\end{array}
\right.
.
\label{eqctrwAVE}
\end{equation}
Clearly the  mean diverges when $\alpha>3$ which is critical for our discussion. Note the peculiarity of the case $\alpha=2$ as the mean
action increment is independent of $v_{{\rm max}}$.

 We now examine the distribution of $s$ in some detail. For  $\alpha>2$ we
 change variables according to $z=v^{\alpha-2} s/c$ in Eq. (\ref{ctrw07})
 and then the lower limit of the integral namely $v=0$ corresponds to $z=0$.
Note that
when $\alpha<2$  the lower limit of integration over $z$ transform to infinity.
We find
\begin{equation}
\phi_1(s) = { c^{1 \over \alpha-2} \over (\alpha-2)v_{{\rm max}}}
s^{ - 1 - {1 \over  \alpha-2} } \gamma\left( { \alpha-1 \over \alpha -2} , {v_{{\rm max}} ^{\alpha-2} s \over c} \right) \  \  \ \mbox{when} \ \ \ \alpha>2.
\label{ctrw08}
\end{equation}
For large $s$ the lower  incomplete Gamma function is a constant equal to the 
corresponding Gamma
function, hence for large $s$ we find the power law tail
 $\phi_1(s) \propto s^{-1 - 1/(\alpha-2)}$
and from here we see again that when $\alpha>3$
 the mean action increment diverges.  
For example for the experimentally relevant case
 $\alpha=4$ we have
\begin{equation}
\phi_1(s) = { c^{1 \over 2} \over 2 v_{{\rm max}}}s^{-3/2} \gamma\left( {3 \over 2}, {v_{{\rm max}} ^2 s\over c} \right) \ \ \ \mbox{for}  \ \ \alpha=4
\label{ctrw08}
\end{equation}
which gives $\phi_1(s) \sim \sqrt{c \pi} / (4  v_{{\rm max}} ) s^{-3/2}$ when $s \to 
\infty$.  

For $\alpha<2$ the integral Eq. (\ref{ctrw07})
can be solved similarly and the solution   can be expressed
in terms of the upper incomplete Gamma function
\begin{equation}
\phi_1(s) = { c^{1 \over \alpha-2} \over v_{{\rm max}}(2-\alpha)} s^{-1-{1 \over \alpha-2} } \Gamma\left( {\alpha -1 \over \alpha -2} , { (v_{{\rm max}})^{\alpha-2} s \over c} \right), \ \ \ \mbox{when} \ \ \  \alpha<2.
\end{equation}
%
%
%
Now for large $s$ the distribution
 $\phi _{1}(s)$ has an exponential cutoff
for all $\alpha $ with $1\leq \alpha \leq 2$, as can be seen from the
asymptotic expansion of the upper incomplete Gamma function $\Gamma
(a,x)\sim x^{a-1}\exp (-x)$ resulting for $1\leq \alpha <2$ in
\[
\phi _{1}(s)\sim \frac{1}{2-\alpha }\frac{1}{s}\exp (-\frac{s}{c}%
v_{{\rm max}}^{\alpha -2}),
\]%
and for $\alpha =2$ we saw already that $\phi _{1}(s)=\frac{1}{c}\exp (-%
\frac{s}{c})$. For small $s$ one gets a finite value $\phi _{1}(s\rightarrow
0)=\frac{v_{0}^{\alpha -2}}{c(\alpha -1)}$ for all $\alpha >1$. The case $\alpha
=1$ is special in this respect, because here $\phi _{1}(s)$ is given by 
\[
\phi _{1}(s)\sim \frac{1}{cv_{{\rm max}}}\Gamma (0,\frac{s}{cv_{{\rm max}}}),
\]
which blows up at 
 small $s$ like
$\phi_1(s) \sim [- \ln(s/(v_{{\rm max}} c)) - \gamma_1 ]/( v_{{\rm max}}c)$
 where
$\gamma_1\simeq 0.577$ is the Euler Mascheroni constant. We see that $\alpha=1$, which marks the transition between 
phases  with and without a finite mean waiting time, 
also exhibits a transition from
a blow up at the origin of $\phi_1(s)$ to the case where the PDF
of increments is a constant at short $s$. 

To summarize, we see that the
critical value  $\alpha=1$ 
marks the  transition from a finite mean  waiting 
to an infinite mean, while $\alpha=3$ marks the transition between
a finite mean  action increment  to a diverging one. 
The critical value $\alpha=3$  is  specific to the observable under study,
namely the energy of the atom. Considering an observable
like $v(t)^q$ and $q\neq 2$
 would yield other values of the  critical exponent.
Still, energy is a  basic concept for thermodynamics,
so we focus on this particular
observable.  
Also $\alpha=2$ marks a transition, from a power law to an asymptotically
 exponentially decaying  distribution of action increments. However, this does not
play an essential role in our investigations of ergodic properties of the process.  Importantly, the joint PDF of $s$ and $\tilde{\tau}$ does not factorise,
and hence we need to treat the correlations between these variables,
which is what is done in the next subsection. 

\begin{figure}\begin{center}
\epsfig{figure=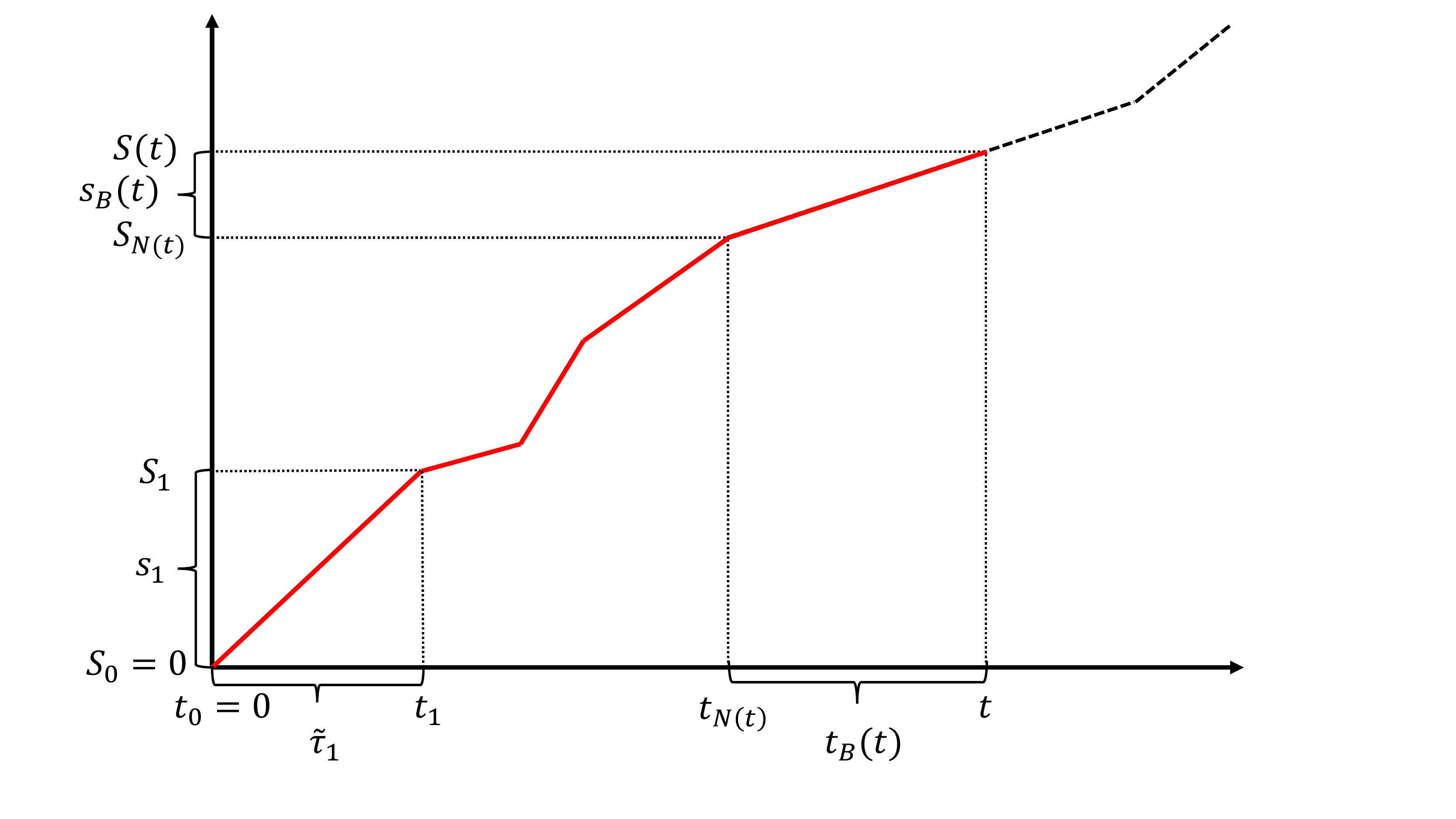, width=0.60\textwidth,trim=0mm 0mm 0mm 0mm, clip}
\end{center}
\caption{
Action-time diagram illustrating the generalized L\'evy walk performed 
by the action ${\cal S}(t)$, Eq. (\ref{ctrw03}). $t_{N(t)}$ and ${\cal S}_{N(t)}$  are the partial 
sums appearing in Eqs. (\ref{eqCONS}) and (\ref{ctrw03}), where $t_B (t)=t-t_{N(t)}$ and $s_B 
(t)=S(t)-S_{N(t)}$ are the backward recurrence time and the associated 
action increment, respectively. For this sample trajectory the number 
$N(t)$ of renewals until time $t$ is $N(t)=4$.
}
\label{fig4a}
\end{figure}

\subsection{Governing Montroll-Weiss
equations for the distribution of the  total action}

 We will now investigate the formalism giving the PDF of the action
${\cal S}(t)$ at time $t$ which is denoted $P({\cal S},t)$. 
In the next subsection we consider some of its long time properties. 
This density is of course normalised according to $\int_0 ^\infty P({\cal S}, t) {\rm d}  {\cal S} =1$.
From the relation $\overline{E}_k(t)= {\cal S}(t) /t$, then from the distribution
of ${\cal S}(t)$ we can gain insights on the ergodic properties of $\overline{E}_k(t)$. 

 Let $Q_N({\cal S},t) {\rm d} t$ be the probability that the $N$-th transition event takes place in the small
interval $(t, t + {\rm d} t)$ and the value of the action
${\cal S}(t)$  switched to ${\cal S}$.  
It is given by the iteration rule
\begin{equation}
Q_{N+1} \left({\cal S} , t\right) = \int_0 ^{{\cal S}} \int_0 ^t 
 {\rm d} s {\rm d} \tilde{\tau}
Q_{N} \left({\cal S} -s, t-\tilde{\tau} \right)\phi_2(s, \tilde{\tau}) 
\label{eqG01} 
\end{equation}
with $Q_0 ({\cal S},t) = \delta( {\cal S}) \delta(t)$ being the initial condition. 
The equation describes the basic property of the process:  to arrive in  ${\cal S}$ at time $t$ when the previous
collision event took place at time $t -\tilde{\tau}$, the previous value of the action was
${\cal S} -s$. Solving this equation is made possible with the  convolution
theorem of the Laplace transform. Let 
\begin{equation}
\hat{Q}_N \left( u, p \right) = \int_0 ^\infty  \int_0 ^\infty 
{\rm d} {\cal S} {\rm d} t
\exp\left( -u  {\cal S}  - p t  \right) 
Q_N \left( {\cal S}, t \right) .
\label{eqG02}
\end{equation}
be  the double Laplace transform  of $Q_n({\cal S},t)$ where $u \leftrightarrow {\cal S}$ and
$p \leftrightarrow t$ are Laplace pairs. Then from the
convolution theorem we have
\begin{equation}
\hat{Q}_N (u, p) = \left[ \hat{\phi}_2 (u,p) \right]^N
\label{eqG03}
\end{equation}
where $\hat{\phi}_2 (u,p)$ is the double Laplace transform of $\phi_2(s,\tilde{\tau})$.
The PDF $P({\cal S},t)$ is in turn given by
\begin{equation}
P\left({\cal S} , t\right) = \sum_{N=0} ^\infty \int_0 ^{\cal S} \int_0 ^t 
{\rm d} s {\rm d} t'_B 
 Q_N\left( {\cal S}-s,t-t'_B\right) \Phi_2 \left( s, t'_B\right).
\label{eqG04}
\end{equation}
Here we sum over the number of events $N$ in the time interval $(0,t)$
and take into consideration that  the measurement time $t$ is in an interval
straddled by   two
collision events, we further integrate over the backward recurrence time $t'_B$.
Finally, the statistical weight function is 
\begin{equation}
\Phi_2 (s, t' _B) = \int_{t' _B} ^\infty {\rm d} \tilde{\tau} \int_0 ^{v_{{\rm max}}} {{\rm d} v \over v_{{\rm max}}}
R(v) \exp \left( - R(v) \tilde{\tau}\right) \delta\left(s- t' _B v^2\right).
\label{eqG06}
\end{equation}
Here the waiting time for the next jump is larger than $t' _B$ since by definition the next transition
takes place at times larger than $t$. In this equation we  average over the speed  i.e. integrate over $v$,
draw the waiting time from an exponential PDF with a velocity dependent rate and
take into consideration the fact that  the last increment
of the action is given by $s=t' _B v^2$ leading to the delta function.

 Now we consider the double Laplace transform of $P({\cal S}, t)$ which is denoted by
$\hat{P}(u,p)$. We again use the convolution theorem, and Eqs. 
(\ref{eqG03},\ref{eqG04}) give, after summing a geometric series,
\begin{equation}
\hat{P} (u,p) = { \hat{\Phi}_2 (u, p) \over 1 - \hat{\phi}_2 (u,p) } . 
\label{eqG07}
\end{equation}
In the context of random walk theory, such a formula is called  the Montroll-Weiss equation \cite{MW},
though typically instead of a double Laplace transform one invokes a Laplace-Fourier transform \cite{KBS,Aghion}. 
In general to invert such an expression
 to the ${\cal S},t$ domain is hard, however,
certain long time limits can be obtained.
In particular from the definition of the
Laplace transform  the mean of the action in Laplace space is 
\begin{equation}
\langle \hat{{\cal S}} (p) \rangle = - {\partial \hat{P} (u,p) \over \partial u} |_{u=0}.
\label{eqG08}
\end{equation}
Hence to get $\langle {\cal S}(t) \rangle$ we need to deal with a single inverse Laplace
transform from $p$ back to $t$. Using Eq. 
(\ref{eqG07}) we find
\begin{equation}
\langle \hat{{\cal S}} (p) \rangle = - {{\partial \hat{\phi_2} (u,p) \over \partial u} |_{u=0}\hat{\Phi}_2 (0, p) 
\over \left[ 1 - \hat{\phi}_2 (0,p) \right]^2 }
- { {\partial \hat{\Phi}_2 \left(u,p\right) \over \partial u} |_{u=0} \over 1 -\hat{\phi}_2 (0,p)}.
\label{eqG09}
\end{equation}
Note that the double Laplace transform  $\hat{\phi}_2(0,p)$ evaluated at
$u=0$ is merely the Laplace $\tilde{\tau} \to p$ transform
of the waiting time PDF $\phi_1(\tilde{\tau})$,  so $\hat{\phi}_2 (0,p)=\hat{\phi}_1(p)$. This of course comes
 from the fact that by integration of 
the joint PDF $\phi_2(s,\tilde{\tau})$ over $s$ we obtain
the marginal PDF
$\phi_1(\tilde{\tau})$.  
The second expression on the right hand side of Eq. (\ref{eqG09})
 is a contribution to the total  action 
from the backward recurrence time, and in the long time
limit and for ordinary processes, with a finite mean waiting time, can be neglected. 
To investigate ergodicity one needs to go beyond the mean and consider the full
distribution of ${\cal S}(t)$, or at least its variance.

\subsection{ 
 Energy is an integrable observable
when $1<\alpha < 3$}

We now obtain $\langle \overline{E}_k(t)\rangle$ using two approaches, the first is based on the Montroll-Weiss machinery Eq. 
(\ref{eqG07}),
 the second exploits the infinite invariant  density. 
When $\alpha<1$ we have a standard steady state since then the mean waiting time is finite. 
We now consider the case $1<\alpha<3$. 
According to Eq. 
(\ref{ctrwaa})
the mean waiting time diverges while
Eq. (\ref{eqctrwAVE}) yields a finite 
 mean action increment $\langle s \rangle$.  Further the observable
$v^2$, representing the energy, is integrable with respect to the infinite density. This is because ${\cal I}_v (v) \propto
v^{ - \alpha}$ for $v \to 0$, and hence the integral $\int_0 ^\infty {\cal I}_v (v) v^2 {\rm d} v$ is finite when $\alpha<3$.

 We are interested in the long time limit of $\langle {\cal S} (t) \rangle$, 
and in Appendix $B$ we derive a rather intuitive equation 
\begin{equation}
\langle {\cal S} (t) \rangle \sim  \langle N (t) \rangle \langle s \rangle
\ \ \ \mbox{and} \ \ \ \ \langle \overline{E}_k (t) \rangle \sim {\langle N(t) \rangle \langle s \rangle \over t}
\label{eq1a3}
\end{equation}
 with  $\langle N (t)\rangle$  the mean number of collisions in the time interval
$(0,t)$. The
mean $\langle N(t) \rangle \propto t^{1/\alpha}$ increases sub-linearly 
because the mean time interval  between collision
events diverges. This can be justified with a hand waving argument as follows. 
When the mean $\langle \tilde{\tau} \rangle$
is finite
we expect from the law of large numbers that
 $\langle N(t) \rangle \sim t/\langle \tilde{\tau} \rangle$. However,
in our case, when $\alpha>1$ the denominator diverges, and is replaced with the effective mean, namely 
$\langle N(t) \rangle \propto t / \int_0 ^t t' \phi_1(t') {\rm d} t' \propto t^{1/\alpha}$. 
Using well known formulas from renewal theory, briefly recapitulated
 in Appendix B, we find
\begin{equation}
\langle {\cal S} (t) \rangle \sim \underbrace{ { c (v_{{\rm max}})^{2 -\alpha} \over 3 - \alpha}}_{\langle s \rangle } 
\underbrace{ { \alpha \sin( \pi/\alpha) \over \pi \Gamma\left( 1 + {1\over \alpha}\right) } \left( { (v_{{\rm max}})^\alpha t \over c } \right)^{1/\alpha}}_{\langle N(t) \rangle}. 
\label{eq1a5}
\end{equation}
This gives the mean of the time averaged  kinetic energy of the particles (mass $m/2=1$)
$\langle \overline{E}_k(t) \rangle \sim   \langle {\cal S}(t) \rangle/  t$.

\subsubsection{Infinite ergodic theory at work}

Infinite ergodic theory makes the calculation of
$\langle E_k(t) \rangle$ or $\langle \overline{E}_k(t) \rangle$ 
 easy in the sense that one needs only the knowledge of  the invariant density
${\cal I}_v(v)$. 
Eqs. (\ref{eq12},
\ref{eqFV},
\ref{eqEVC}) give
\begin{equation}
{\cal I}_v (v) =
\left\{
\begin{array}{c c}
{ 
\sin \pi \gamma \over \pi\Gamma(1+ \gamma)} c^{1 - \gamma} v^{-1/\gamma} \ & \ 0<v<v_{{\rm max}}\\
0 \ & \ \mbox{otherwise,}
\end{array}
\right. 
\label{eqINF}
\end{equation}
which is clearly non-normalisable. 
Then the ensemble average 
Eq. (\ref{eqave02})
\begin{equation}
\boxed{
\langle E_k (t) \rangle = \langle v^2 (t) \rangle\sim
 { \int_0 ^{v_{{\rm max}}} v^2 I_{v}(v) {\rm d} v \over t^{1-\gamma} }=
{\sin (\pi \gamma) \over \Gamma(1+\gamma)  \pi } 
{(v_{{\rm max}})^{3 -1/\gamma}  \over 3 -1/\gamma} 
 \left({ c\over t}\right)^{1-\gamma} 
}
\label{eqINF1}
\end{equation} 
and Eq.
(\ref{eqave07})
 gives $\langle \overline{E}_k (t) \rangle= \langle E_k(t) \rangle /\gamma$.
On the other hand $\langle \overline{E}_k(t) \rangle = \langle {\cal S}(t) \rangle /t$ and as expected Eq.
(\ref{eq1a5}) divided by $\gamma$  gives the same results as in
Eq. (\ref{eqINF1}). The calculation using  infinite
ergodic theory is  straight forward and is essentially
similar to the averaging we perform  with ordinary equilibrium calculations and in that sense it provides a tool more convenient  compared
to the Montroll-Weiss random walk approach. Of course, the latter yields insights since
it connects the  mean of the observable to the number of renewals
$\langle N(t) \rangle$. 
Note that as $\gamma \to 1/3$ Eq. 
(\ref{eqINF1}) exhibits a blow up, further $\lim_{\gamma \to 1/3 } v^{3 - 1/\gamma} = 1$ and in this sense the energy is switching to a behaviour that is 
$v_{{\rm max}}~$ independent, more generally it is independent of the parent velocity PDF
$f(v)$. This marks the transition to the phase
where the energy is no longer an integrable observable, as discussed in Sec.
\ref{SecNI} below. 

\begin{figure}\begin{center}
\epsfig{figure=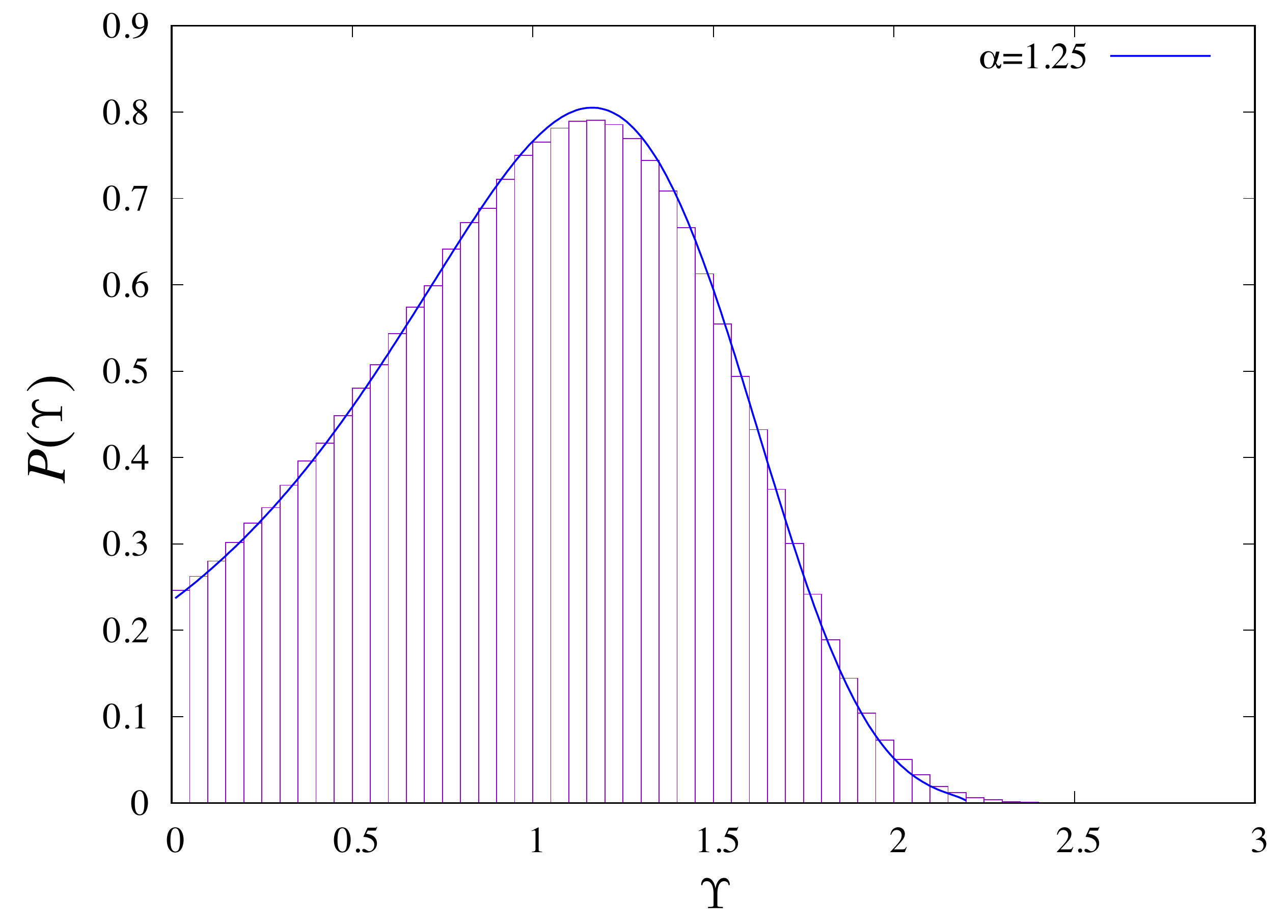, totalheight=0.30\textheight, width=0.55\textwidth,trim=0mm 0mm 0mm 0mm, clip}
\end{center}
\caption{
To investigate the ergodic properties of the system,
we simulate the cooling process model with a uniform velocity distribution
$f(v)$ with maximum velocity unity $v_{{\rm max}}=1$,
 and and the rate  $R= v^\alpha$ so $c=1$.
 We then follow
 velocity paths as a function of time  and
from each path we obtain the time the particle  spent 
in an
interval $v_a<v<v_b$. Namely, our 
 observable is the indicator function. For standard ergodic processes, e.g. for the velocity of a gas particle obeying Maxwell statistics, this time divided by  a long measurement time is 
approaching the  probability of being in the mentioned velocity domain. For the laser cooling  model, this time average fluctuates, the Mittag-Leffler distribution describes the corresponding statistics. 
Specifically we show the histogram of the  normalised random 
variable $\Upsilon= \int_0 ^t I(v_a<v(t') <v_b) {\rm d} t' / \langle \int_0 ^t I(v_a< v(t') <v_b) {\rm d} t'\rangle$
which perfectly matches the theory Eq. 
(\ref{eqML}). 
 Here $\alpha=1.25$ so
 $\gamma=0.8$, namely we are
not too far from the ergodic phase $(\gamma=1)$, hence we see a peak in the distribution
of $\Upsilon$
close to its mean which is unity. 
For ergodic process  we will find a delta peak centred on unity and
this can be found in our model by choosing $\alpha<1$. We used $t=10^9$, $v_a=0.4$ and $v_b=1$.
}
\label{fig4}
\end{figure}

\section{Fluctuations described by the Darling-Kac theorem}

 We now consider the fluctuations of the time average of the energy
for $1<\alpha<3$. 
Since $\overline{E_k}= {\cal S}(t) /t$, we investigate the distribution
of ${\cal S}(t)$ for long times namely $P({\cal S},t)$,  using its double
Laplace transform 
Eq. (\ref{eqG07}). From  Eqs. 
(\ref{eqFV},
\ref{eqEVC},\ref{ctrw04}) and  the definition of Laplace transform
\begin{equation} 
\hat{\phi}_2 (u,p) = \int_0 ^\infty {\rm d} s \int_0 ^\infty {\rm d} \tilde{\tau} \exp( - u s - p \tilde{\tau}) {1 \over v_{{\rm max}}} \int_0 ^{v_{{\rm max}}} {\rm d} v R(v)
\exp[ - R(v) \tilde{\tau}] \delta (s - v^2 \tilde{\tau}).
\label{eqFLUC01}
\end{equation}
Integrating over $s$ and $\tilde{\tau}$ gives
\begin{equation}
\hat{\phi}_2 (u, p ) = {1 \over v_{{\rm max}}} \int_0 ^{v_{{\rm max}}} {R(v) \over p + R(v) + u v^2} {\rm d} v. 
\label{eqFLUC02}
\end{equation}
Similarly we find
\begin{equation} 
\hat{\Phi}_2 (u,p) = {1 \over v_{{\rm max}}} \int_{0} ^{v_{{\rm max}}} {1  \over p + R(v) + u v^2} {\rm d} v . 
\label{eqFLUC03}
\end{equation}
For $u=0$ we have $\hat{\Phi}_2(0,p) = [ 1 -  \hat{\phi}_2 (0,p)]/p$,
which is the Laplace transform of the probability of not making a jump,
also called the survival probability. 

Since $t$ is large and so is $S$ we consider the limit $u \to 0$ and
$p\to 0$. This limit is considered under the condition that
the ratio 
$p^\gamma /u \rightarrow \mbox{const}$, or ${\cal S} / t^\gamma$,
remains finite. 
This  scaling is anticipated  from the
 behaviour of the moments of ${\cal S}$
for example Eq.  
(\ref{eq1a5}).  

We first consider the numerator of Eq. 
(\ref{eqG07}).
Changing variables in Eq. 
(\ref{eqFLUC03})
according to $z^\alpha=  R(v)/p=
v^\alpha / (p c)$ we find
\begin{equation}
\hat{\Phi}_2 (u,p) =
{ 1 \over v_{{\rm max}} p } \int_0 ^{v_{{\rm max}}/ (p c)^\gamma } { ( p c)^\gamma \over 1 + z^{1/\gamma} +  (p c)^{2 \gamma} u z^2/p } {\rm d} z.
\label{eqFLUC04}
\end{equation}
As communicated we consider the case  $1/3<\gamma<1$ and in this regime
$p^{2 \gamma-1} u = p^{3 \gamma-1} (u/p^\gamma) \rightarrow 0$
since the ratio  $u/p^\gamma$ is fixed.  
We then take the upper limit of integration to infinity, and hence to leading order
\begin{equation}
\hat{\Phi}_2 (u,p) \sim { c^\gamma p^{\gamma-1} \over v_{{\rm max}}} \int_0 ^\infty {{\rm d} z \over 1 + z^{1/\gamma} }
\label{eqFLUC05}
\end{equation}
and $\int_0 ^\infty (1 + z^\gamma)^{-1} {\rm d } z = \gamma \pi \csc (\gamma \pi)$ with $\gamma<1$. 
 
We now  consider the denominator of Eq. (\ref{eqG07})
 and  note that $\lim_{u,p\to 0} \hat{\phi}_2(u,p) =1$ from normalization. 
Rewriting we find
\begin{equation}
1 - \hat{\phi}_2 (u,p) = 
{1 \over v_{{\rm max}}} \int_0 ^{v_{{\rm max}}} \left[ 1 - { R(v) \over p + R(v) + u v^2} \right] 
{\rm d} v
=
\underbrace{{p \over v_{{\rm max}}}  \int_0 ^{v_{{\rm max}}} { {\rm d}v \over p + R(v) + u v^2}}_{G_1}
+ 
\underbrace{ {u \over v_{{\rm max}}}  \int_0 ^{v_{{\rm max}}} {v^2 {\rm d} v \over p + R(v) + u v^2}}_{G_2}. 
\label{eqFLUC06}
\end{equation}
In the limit  $G_1 \sim (c^\gamma p^\gamma/v_{{\rm max}}) \int_0 ^\infty {\rm d} z(1 + z^\gamma)^{-1}$ and $G_2\sim  c (v_{{\rm max}})^{2-\alpha} u /(3 -\alpha)$, or using Eq.
(\ref{eqctrwAVE})
$G_2 \sim u \langle s \rangle$. We see that
$G_1\propto p^\gamma$ and $G_2\propto u$ so these two terms are of the same order. 

Inserting Eqs. 
(\ref{eqFLUC05},
\ref{eqFLUC06})
in Eq. 
(\ref{eqG07})
we find
\begin{equation} 
\hat{P} (u,p) \sim {  ({\cal T}_1)^\gamma  p^{\gamma -1} \over ({\cal T}_1 p)^\gamma +\langle s \rangle u}
\label{eqFLUC07}
\end{equation}
with $({\cal T}_1 )^\gamma = \gamma \pi  c^\gamma / (v_{{\rm max}} \sin \gamma \pi) $.
Now the inverse Laplace transform $u\rightarrow {\cal S}$ gives
\begin{equation}
\hat{P}({\cal S}, p ) \sim {({\cal T}_1)^\gamma p^{\gamma -1} \over \langle s \rangle} \exp\left[ - { ({\cal T}_1 p)^\gamma {\cal S} \over \langle s \rangle} \right]=
- { {\rm d} \over {\rm d} p } {1 \over \gamma {\cal S} } \exp \left[
- { ( {\cal T}_1 p )^\gamma {\cal S} \over \langle s \rangle } \right].
\label{eqFLUC08}
\end{equation}
Let $l_{\gamma,1}(t)$ be the one-sided L\'evy stable
distribution with index $\gamma$ such that $l_{\gamma,1}(t) \leftrightarrow
\exp( - p^\gamma)$ are Laplace pairs. Using $-d/dp \leftrightarrow t$
we obtain the PDF of the action for $1/3<\gamma<1$
sometimes  called after Mittag-Leffler 
\begin{equation}
P \left( {\cal S} , t \right) \sim 
{1 \over \langle s \rangle} {\tilde{t} \over \gamma \tilde{{\cal S}}^{1 + 1/\gamma}} 
l_{\gamma,1} \left( {\tilde{t} \over \tilde{{\cal S}}^{1/\gamma}} \right)
\label{eqFLUC09}
\end{equation}
with $\tilde{t} = t/{\cal T}_1$ and  $\tilde{{\cal S}} = {\cal S} / \langle s \rangle$. The one-sided L\'evy stable distribution is well documented \cite{Gorska}
for example
in Mathematica  and hence
it is easy to plot this solution. More importantly, this distribution shows
exactly how the fluctuations of the time averaged energy behave.

 It is the custom to consider normalised random
variables with  unit mean, namely instead of ${\cal S}(t)$ 
we investigate $\Upsilon= {\cal S}(t) / \langle {\cal S}(t) \rangle$. 
This by definition is also $\Upsilon= {\overline E}_k(t) /\langle \overline{E}_k (t) \rangle$ and using  
Eq. (\ref{eqave07}) $\Upsilon= \gamma \overline{E}_k(t) /\langle E_k(t) \rangle$
and importantly  the denominator can be obtained by a simple phase space
average of $v^2$ employing the infinite density. 
 Then asymptotically for large
$t$,  the PDF of $\Upsilon$
is time invariant and
according 
to Eq. (\ref{eqFLUC09}) given by the Mittag-Leffler law
\begin{equation}
\boxed{
\mbox{ML} \left( \Upsilon \right) = { [\Gamma \left( 1 + \gamma \right)]^{1/\gamma} \over \gamma \Upsilon^{1 + 1/\gamma} } 
l_{\gamma,1} \left( { [\Gamma (1 + {1 \over \gamma} )]^{1/\gamma}  \over \Upsilon^{1/\gamma}} \right).
}
\label{eqML}
\end{equation}
While not proven here, the Darling-Kac theorem states that this result is valid for any observable integrable with respect to the infinite measure,
namely the PDF of  $\Upsilon= \overline{{\cal O} }(t) /\langle {\overline O} (t) \rangle= \gamma \overline{{\cal O}}(t)  / \langle {\cal O} (t)\rangle$ is
asymptotically given by  $\mbox{ML} (\Upsilon)$. 
 We demonstrated this universality
with a few  examples, choosing as observable both the kinetic energy
and  the indicator function,
see Figs. 
\ref{fig4}-\ref{fig6}. 
In the limit $\gamma \to 1$ the PDF $\mbox{ML}(\Upsilon)$ 
approaches a delta function, corresponding to Birkhoff's ergodic theorem,
while in the opposite limit $\gamma \to 0$ the PDF $\mbox{ML} (\Upsilon)$
is exponential with mean unity, see Figs. \ref{fig4} - \ref{fig6} which explore
this trend. It should be recalled that the limit $\gamma\to 0$ is valid here only for the indicator function, and not for the energy, since the former is integrable in the whole domain $0<\gamma<1$ (if $v_a>0$), while the latter only in
the interval $1/3<\gamma<1$.  
Mathematically the Mittag-Leffler distribution holds in the long time
limit  for
 $\Upsilon = N(t)/\langle N(t) \rangle$ where
$N(t)$ is the random  number of collisions/renewals in the process
until time $t$. Roughly speaking when $\langle s \rangle$ is finite we
have ${\cal S}(t) \simeq N(t)  \langle s \rangle$, and hence the
statistics of the time averaged kinetic energy
 and ${\cal S}(t)$ are given by the same law as the statistics of $N(t)$.  What is remarkable is that a similar behaviour  holds for any integrable
observable and that the mean of the observable is easy to compute
with  the infinite
density.

 It is easy to find the long time limit of the moments of the process
using Eq. (\ref{eqFLUC08}) and then inverting to the time domain.
For example $\langle \hat{S}(p) \rangle \sim \langle s \rangle p^{-\gamma-1} / ({\cal T}_1)^\gamma$ for small $p$, which gives Eq. 
(\ref{eq1a5}). Similarly the $\mbox{EB}$ parameter characterising the fluctuations
of the time average \cite{PRLctrw} in the long time limit is given by
\begin{equation}
\boxed{
\mbox{EB} = { \langle \overline{E_k}^2 (t) \rangle - \langle \overline{E_k}(t)\rangle^2 \over
\langle \overline{E_k}(t) \rangle^2} = 
{ \langle {\cal S}^2(t)  \rangle - \langle {\cal S}(t)  \rangle^2 \over \langle {\cal S} (t) \rangle ^2 } = 
{ 2 \Gamma^2(1+\gamma) - \Gamma(1 + 2 \gamma) \over \Gamma(1 + 2 \gamma)}.
}
\label{eqFLUC10}
\end{equation}
Thus when $\gamma=1$ the fluctuations vanish, since then we enter the ergodic phase, when the mean trapping time is finite. Recall that for the energy observable,  this equation is
valid for $\gamma>1/3$ since we assumed that $\langle s \rangle$ is finite.

\begin{figure}\begin{center}
\epsfig{figure=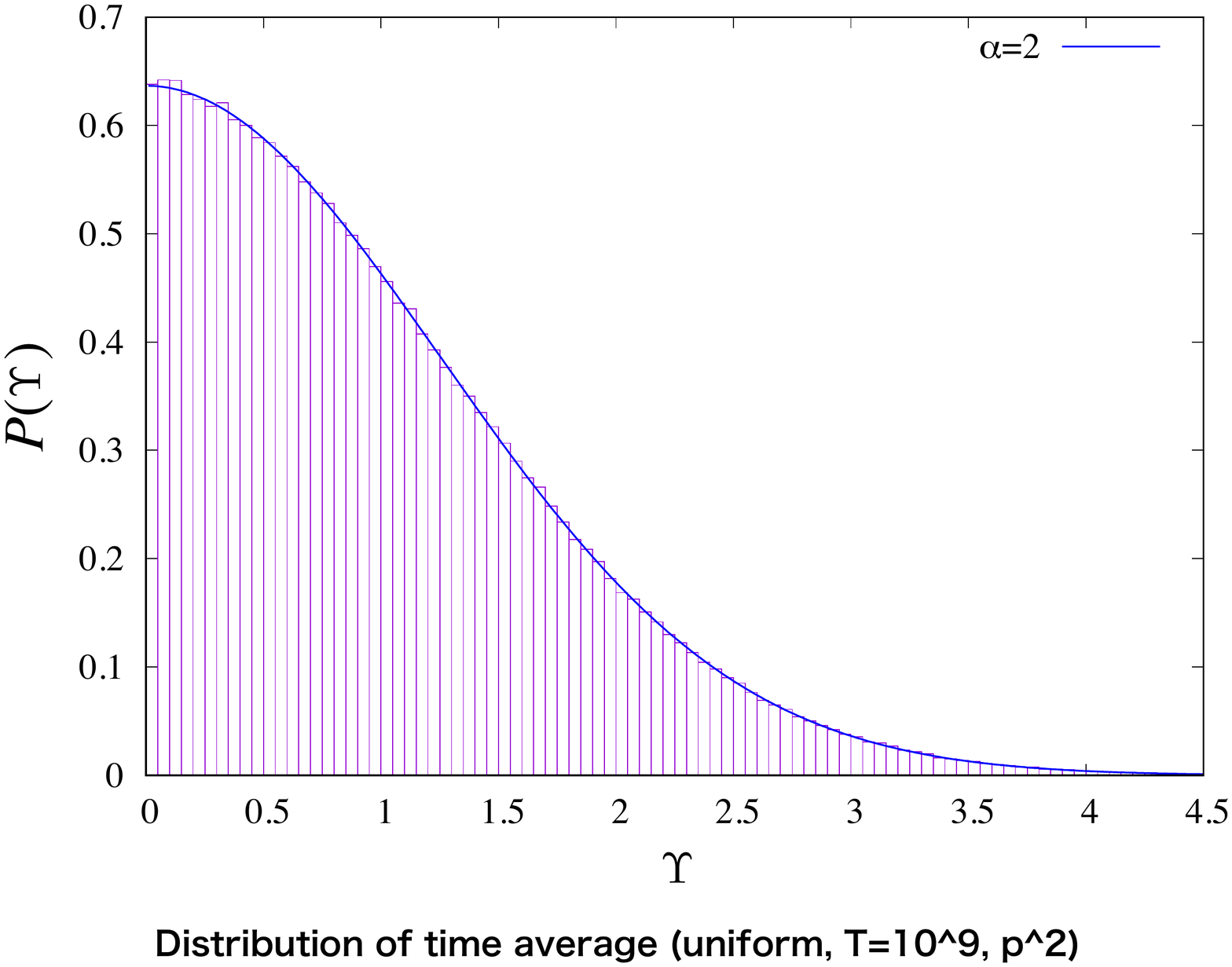, totalheight=0.25\textheight, width=0.45\textwidth,trim=0mm 0mm 0mm 0mm, clip}
\end{center}
\caption{
Same as Fig. \ref{fig4} however now the observable is the kinetic energy $v^2(t)$ and $\alpha=2$. For this choice of $\alpha$ the kinetic energy is integrable with respect to the infinite density, and hence fluctuations of its
time average obey the Darling-Kac law. We present $\mbox{ML} (\Upsilon)$  PDF of the normalised 
time average $\Upsilon=\int_0 ^t v^2(t') {\rm d} t'/\langle \int_0 ^t v^2(t') {\rm d} t' \rangle$, Eq.
(\ref{eqML}).
Since $\gamma=1/2$ the Mittag-Leffler PDF is half a Gaussian presented as the
theory in the figure.  
We used $10^6$ particles and the measurement time was $t= 10^9$. 
}
\label{fig5}
\end{figure}

\begin{figure}\begin{center}
\epsfig{figure=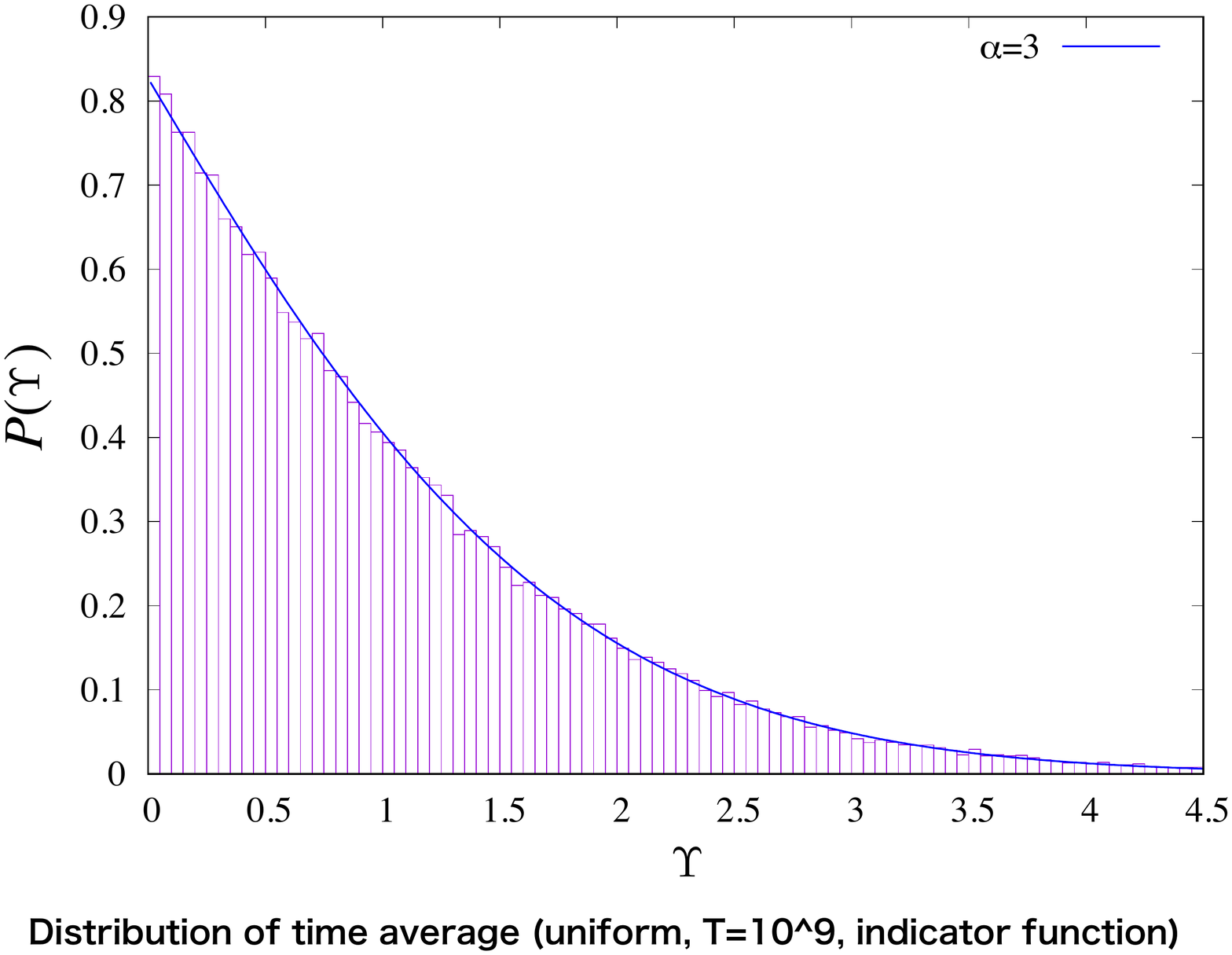, totalheight=0.25\textheight, width=0.45\textwidth,trim=0mm 0mm 0mm 0mm, clip}
\end{center}
\caption{
Distribution of $\Upsilon$ where the later is the
the time spent by a path in a domain $(v_a,v_b)$ over its mean (same as in
Fig. \ref{fig4}) however now $\alpha=3$.
Like any other observable integrable with respect to the infinite density
this normalised  functional of the path obeys Mittag-Leffler statistics Eq. (\ref{eqML}). Since $\gamma=1/3$ the peak of
the histogram of $\Upsilon$ is at the origin, unlike the case presented
in Fig. \ref{fig4}. In that sense the fluctuations of time averages
are larger when $\gamma$ is small compared to $\gamma$ close to unity 
as expected.  
When $\gamma\to 0$ the distribution of $\Upsilon$ for this integrable
observable is exponential and
then the corresponding $\mbox{EB}$ parameter attains its  largest value $\mbox{EB} =1$. 
}
\label{fig6}
\end{figure}

\section{Kinetic energy- the  non-integrable phase $3<\alpha$}
\label{SecNI}

 When considering observables like the kinetic energy we have three
types of behaviours
\begin{equation}
\boxed{
\langle E_k(t) \rangle \sim  \left\{
\begin{array}{l l}
\int_{0} ^{\infty} v^{2} \rho^{{\rm eq}} (v) {\rm  d} v \ & \ \  1<\gamma, \\
\ & \ \\
{ \int_{0} ^{\infty} v^2 {\cal I}_v (v) {\rm d} v \over t^{1-\gamma}  } \ & \  \
1 / 3  < \gamma < 1, \\
\ & \ \\
{ \int_{0} ^{\infty} x^2 g(x) {\rm d} x \over t^{2 \gamma}}  \ & \ \ 
0< \gamma<1/3
\end{array}
\right.
}
\label{eqNI01}
\end{equation}
where we used 
Eqs. 
(\ref{eq05},
\ref{eqave02}).
This, as mentioned,
corresponds to cases where the mean waiting time is finite ($1<\gamma$), the mean time
 diverges but the mean action increment  $\langle s \rangle$ is finite 
($1/3<\gamma<1$),
 and finally 
the case where both diverge $(0<\gamma<1/3)$. 
In the first case standard ergodic theory holds, in the second 
 the Darling-Kac theorem is valid for the energy observable, 
finally we have a phase
where the kinetic energy is non-integrable with respect to the infinite
density  $0<\gamma<1/3$, and this is the  case  which is  now
treated. Notice that both $g(x)$ and $\rho^{{\rm eq}}(x)$ are perfectly normalisable distributions, so the intermediate phase $1/3 < \gamma<1$  is in that sense unique. 

\begin{figure}\begin{center}
\epsfig{figure=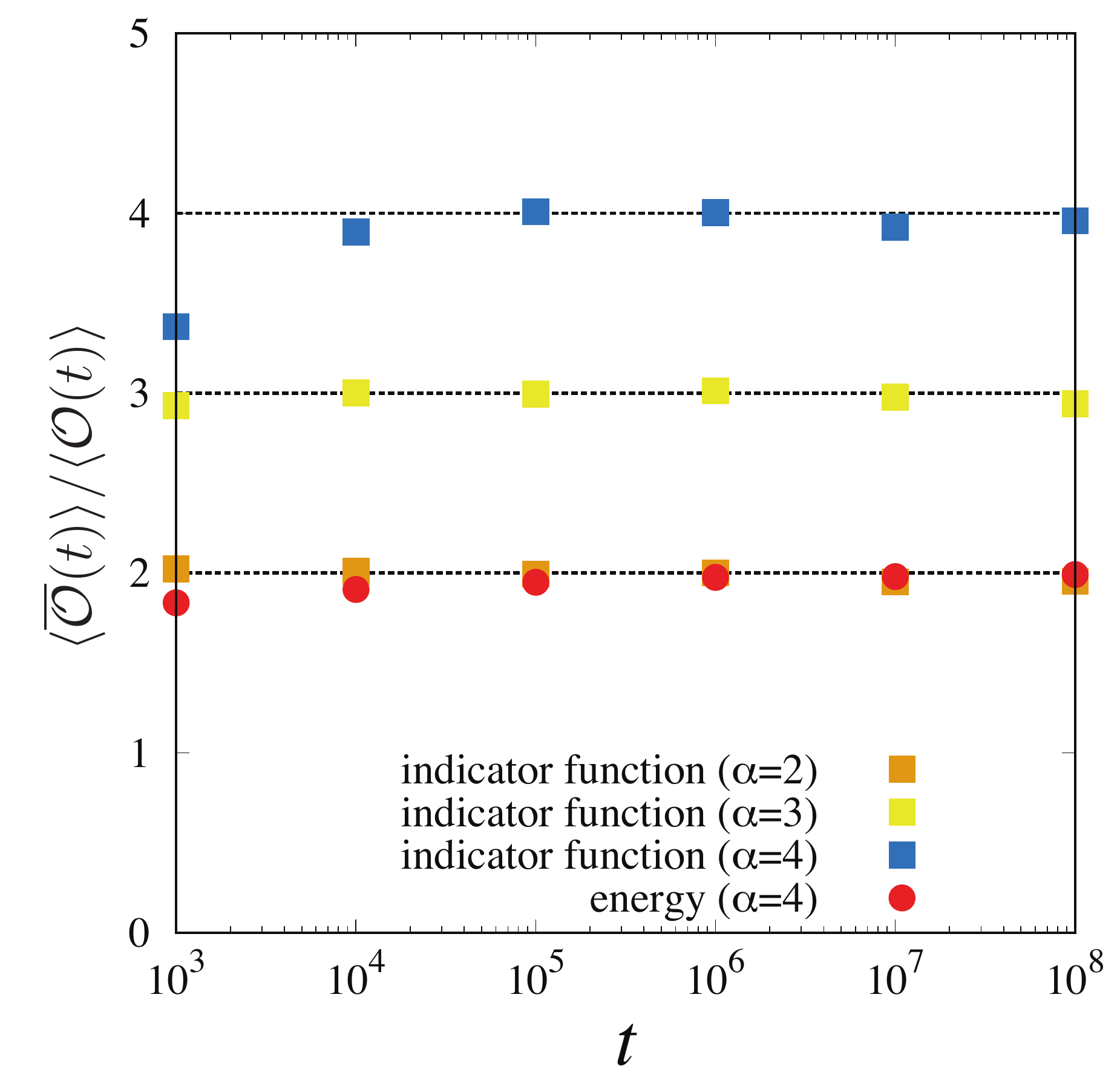, width=0.45\textwidth,trim=0mm 0mm 0mm 0mm, clip}
\end{center}
\caption{
We show the dimensionless ratio $\langle \overline{{\cal O}} \rangle/\langle {\cal O} \rangle$ obtained from numerical simulations versus time.
In the long time limit and if the observable is integrable with respect
to the infinite density, we have according to Eq. 
(\ref{eqave07}) 
$\langle \overline{{\cal O}} \rangle/\langle {\cal O} \rangle=\alpha$, a behavior we illustrate here for the indicator function, which as mentioned in the
text, is an integrable  observable for any choice of $\alpha$. The energy observable, is non integrable for $\alpha=4$.
 Then theory
Eq. (\ref{eqNI04})
 predicts 
$\langle \overline{{\cal O}} \rangle/\langle {\cal O} \rangle=1/(1-2/\alpha)=2$, nicely matching the numerics. 
}
\label{fig6}
\end{figure}

We did not present the derivation of Eq. (\ref{eqNI01}) for
$0<\gamma<1/3$  since the result is rather intuitive.
It means that in this regime the contribution to the  kinetic
energy comes from the slow atoms where the scaling solution is valid. 
Technically we  consider $\int_0 ^\infty v^2 \rho(v,t) {\rm d} v$,
 and divide the integration into two parts, in the
first  the density $\rho(v,t)$ in the small $v$  inner region 
is the scaling solution and in the second
the density is approximated by the  outer solution, namely
the  infinite density.
Then 
after integrating over the observable $v^2$ we can show
 that in
the long time limit  the former is the leading term. 
For the case $1/3<\gamma<1$ the observable is integrable with respect to the
infinite measure  and
then the opposite situation is found. When $\gamma=1/3$ or $\gamma=1$ one
finds logarithmic corrections not treated here.

 Using Eq. 
(\ref{eq11})
we find
\begin{equation}
\langle E_k(t) \rangle = \langle v^2(t) \rangle \sim
\underbrace{\int_{0} ^{\infty}  x e^{- x^{1/\gamma} } \int_{0} ^{x} e^{ z^{1/\gamma} } {\rm d} z {\rm d} x }_{J_\gamma}  {\sin( \pi \gamma) \over \Gamma(1 + \gamma) \pi \gamma} \left( { c \over t} \right)^ {2 \gamma} \ \ \ \mbox{for} \ \ 0<\gamma<{1 \over 3}.
\label{eqNI02}
\end{equation}
Here the kinetic energy is independent of the specific details of 
$R(v)$ besides $c$ and $\gamma=1/\alpha$, namely the parameters
controlling the behaviour of the rate of escape from the trap
at small $v$.
This is very different for the case $ 1/3<\gamma$.
Further according to Eq. (\ref{eqNI02}) 
the velocity PDF $f(v)$ is not at all influencing
the long time dynamics of the mean kinetic energy, though we assume that
this distribution has finite moments, and support on zero velocity. 

%
%

The integral $J_{\gamma }$ 
in Eq. (\ref{eqNI02}),
 can be evaluated analogously to the
calculation of the normalization of $g(x)$ in Eq. 
(\ref{eq11a}).
The integral converges only for $\gamma
<1/3$, and  one obtains 
\begin{equation}
J_{\gamma }=\frac{\pi \gamma }{\sin (3\pi \gamma )}\frac{\Gamma (1+\gamma )}{%
\Gamma (1-2\gamma )},  \label{Eq.(68)}
\end{equation}
where we can see explicitly the divergence at $\gamma =1/3$ and that $%
\lim_{\gamma \rightarrow 0}J_{\gamma }\rightarrow 1/3$. Inserting this
into 
(\ref{eqNI02}),
 we get for the ensemble averaged kinetic energy
\begin{equation}
\left\langle E_{k}(t)\right\rangle \sim \frac{\sin (\pi \gamma )}{\sin (3\pi
\gamma )}\frac{1}{\Gamma (1-2\gamma )}c^{2\gamma }t^{-2\gamma }.
\label{EA E_k}
\end{equation}
From here, we obtain the expectation of  the time average  immediately
\begin{equation}
\left\langle \overline{E_{k}(t)}\right\rangle \sim \frac{\left\langle
E_{k}(t)\right\rangle }{1-2\gamma }\sim \frac{\sin (\pi \gamma )}{\sin (3\pi
\gamma )}\frac{1}{\Gamma (2-2\gamma )}c^{2\gamma }t^{-2\gamma }.
\label{TEA E_k}
\end{equation}
Eqs. (\ref{EA E_k}) and (\ref{TEA E_k}) confirm that in this phase, in
contrast to the phases with $\gamma >1/3$, the average energy is independent
of the details of the parent velocity distribution $f(v)$, especially here
it is independent of $v_{\max }$.

The asymptotic relations between the ensemble averaged kinetic energy $%
\left\langle E_{k}(t)\right\rangle $ and its time average $\left\langle 
\overline{E_{k}(t)}\right\rangle $ in the various phases can therefore be
summarized as
\begin{equation}
\boxed{
\langle \overline{E}_k(t) \rangle \sim \left\{
\begin{array}{l l}
\langle E_k \rangle   \ & \ \ 1< \gamma, \\
\ & \ \\
{\langle E_k(t) \rangle \over \gamma}    \ & \ \ 1/3< \gamma<1, \\
\ & \ \\
{\langle E_k(t) \rangle \over 1-2 \gamma}    \ & \ \ 0<\gamma<1/3.
\end{array}
\right.
}
\label{eqNI04}
\end{equation}

 To obtain the fluctuations of the time averaged energy, namely to calculate the $\mbox{EB}$
 parameter Eq.
(\ref{eqFLUC10}) further work is required. We need to evaluate
the second moment of the action which in Laplace space is
$\langle \hat{{\cal S}}^2(p) \rangle = { \partial^2 \hat{P}(u,p) \over \partial u^2}|_{u=0}$ and then 
using Eq.
(\ref{eqG07})
\begin{equation}
\langle \hat{{\cal S}}^2 (p) \rangle= \underbrace{{ \Phi_2''(0,p) \over 1 - \phi_2(0,p)}}_{I_1(p)}+
\underbrace{{2 \Phi_2 ' (0,0) \phi_2'(0,p) \over [ 1 - \phi_2(0,p)]^2}}_{I_2(p)} +
\underbrace{ {\Phi_2 (0,p) \phi_2 '' (0,p) \over [ 1 -\phi_2(0,p)]^2}}_{I_3(p)} +
\underbrace{ {2 \Phi_2 (0,p) [\phi_2 ' (0,p)]^2 \over [1 - \phi_2(0,p)]^3 }}_{I_4(p)}.
\label{eqEBp01}
\end{equation}
Unfortunately all four terms are contributing to the small $p$ limit under investigation.
This means that unlike the case $1/3<\gamma$, here  the effect of the last interval
in the sequence, which we called the backward  recurrence time, is dominating the statistics
of the time average of the energy. 
The details of the calculations are presented in the subsection below.
We find asymptotically for $0<\gamma<1/3$ 
\begin{equation}
\boxed{
\mbox{EB} = { \langle \overline{E_k}^2(t) \rangle - \langle \overline{E_k}(t)\rangle^2 \over
\langle \overline{E_k}(t) \rangle^2} = 
{ \langle {\cal S}^2(t)   \rangle - \langle {\cal S}(t)  \rangle^2 \over \langle {\cal S} (t) \rangle ^2 } = 
{ 2 \Gamma^2(2 - 2 \gamma) \over \Gamma(3 - 4 \gamma) } \left[ {
\sin^2(3 \pi \gamma) (1 - 5 \gamma) \over \sin (\pi \gamma) \sin (5 \pi \gamma)} + 3\gamma \right] - 1.
}
\label{eqNI13}
\end{equation}
This expression clearly differs from the $\mbox{EB}$ parameter found in the Darling-Kac phase $1/3<\gamma$,
Eq. 
(\ref{eqFLUC10}).
The latter is universal in the sense that it is valid for any integrable observable. In contrast here
the fluctuations are specific to a non-integrable  observable, namely  the energy
provided that  $0<\gamma<1/3$. 
Note that when $\gamma\to 1/3$ from below and above
$\mbox{EB}= 2 \Gamma^2(4/3)/\Gamma(5/3) -1\simeq 0.766$, namely Eq.  (\ref{eqFLUC10})
and Eq. (\ref{eqNI13}) match, so the $\mbox{EB}$ parameter is a continuous function of $\gamma$
while
its derivative is not. In the range $0<\gamma<1/3$ the $\mbox{EB}$ parameter has a minimum,
an effect that we cannot explain intuitively. 
In Fig. 
\ref{fig7} we present numerical results for the $\mbox{EB}$ parameter versus $0<\gamma<1$, comparing it to the analytical theory.  
A clear transition is observed when the energy observable switches from an integral to a non-integrable
observable, a transition found when $\gamma=1/3$.

In the limit $\gamma\to 0$ we find from Eq.
(\ref{eqNI13})
 $\mbox{EB}=4/5$. To understand this limit we notice that
\begin{equation}
\lim_{\gamma \to 0} \mbox{EB}=
{ \langle  (E_k)^2 \rangle - \langle E_k \rangle^2 \over \langle E_k \rangle^2}
\label{eqGAMMA0}
\end{equation}
where unlike the definition in  Eq. (\ref{eqNI13}) here we have the ensemble
averaged energy not the time averages. In this limit  we have a stagnation effect in the sense that in the measurement time $t$ the system remains
in  a particular though random velocity state for practically the whole duration of the process
namely $\overline{E_k}=  \int_0 ^t   E_k(t) {\rm d} t/t = v^2 t/t=v^2$.  
Note that we consider here the limit where $t$ is made long and only then $\gamma\to 0$. It is easy to find $\langle E_k \rangle = \langle v^2 \rangle$ and $\langle (E_k)^2 \rangle = \langle v^4 \rangle$ in the limit $\gamma\to 0$ using
Eqs. 
(\ref{eq11},
\ref{eq11a}). In particular $\exp( - x^{1/\gamma}/c) = 1$ if $0<x<1$ otherwise it is zero since $\gamma\to 0$. Hence using Eq. (\ref{eq11})
$\langle v^2(t) \rangle = \int_0 ^1 x \int_0 ^x {\rm d} z {\rm d} x (c/t)^{2\gamma}= (1/3) (c/t)^{2 \gamma}$ and similarly
$\langle v^4(t)\rangle = (1/5) (c/t)^{4 \gamma}$ and indeed
$\lim_{\gamma\to 0} \mbox{EB} = [(1/5)-(1/3)^2]/(1/3)^2=4/5$.
To conclude Eq. 
(\ref{eqNI13}) makes perfect sense in the limit $\gamma\to 0$ marking the stagnation of the dynamics, and $\gamma \to 1/3$ marking the transition to
the Darling-Kac phase, in between the solution is not trivial.

\begin{figure}\begin{center}
\epsfig{figure=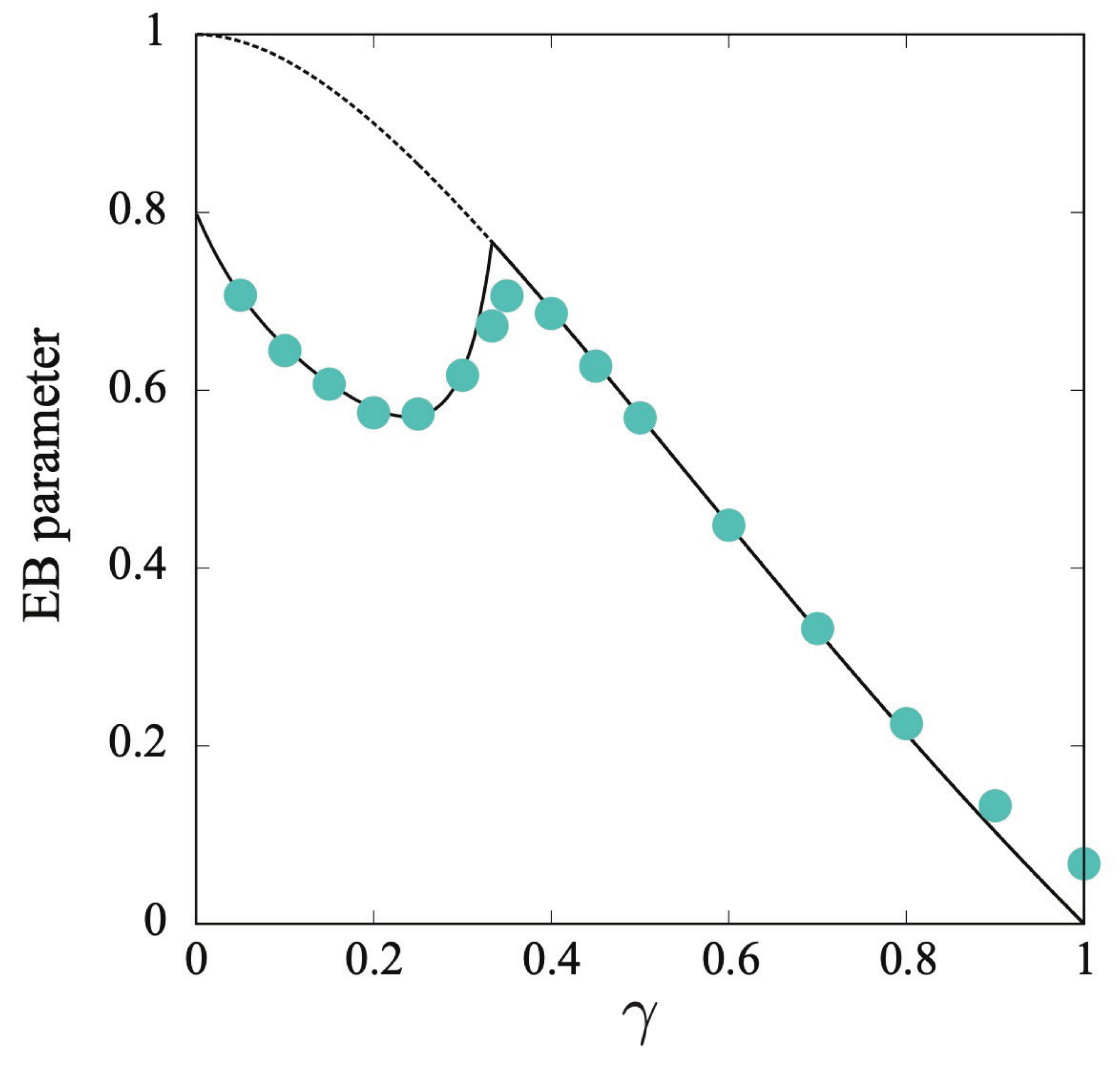, width=0.45\textwidth,trim=0mm 0mm 0mm 0mm, clip}
\end{center}
\caption{
The $\mbox{EB}$ parameter versus $\gamma$ for the kinetic energy observable $v^2$.
When $\mbox{EB}=0$ we have an ergodic phase found when $\gamma>1$. 
This ergodicity breaking parameter is a measure for the fluctuations 
of the time averages  and for 
$1/3<\gamma<1$ is given by Eq.  
(\ref{eqFLUC10}).
That formula is  valid when ever 
the observable is integrable with respect to the infinite measure.
For $0<\gamma<1/3$ the kinetic energy
observable is non-integrable, and 
the system enters a different phase, described by Eq. 
(\ref{eqNI13}).
The dots are obtained from finite time simulations
of a model with a parent velocity distribution $f(v)$ which is uniform and
the number of particles used was $10^5$.  
The continuous dotted curve is Eq. 
(\ref{eqFLUC10}) plotted in the full range $0<\gamma<1$. It  holds for example for  the indicator function which
is an integrable observable in this range, and hence this observable does  not exhibit the discontinuous
behaviour unlike the energy observable. 
}
\label{fig7}
\end{figure}

\subsection{Formula for $\langle \overline{E}_k \rangle$}

We now derive  Eq. 
(\ref{TEA E_k})
the reader not interested in this technicality 
may  of course skip this sub-section.
Specifically we use as a model the uniform  velocity PDF $f(v)$ Eq. 
(\ref{eqFV}) and the  rate function $R(v)$ 
Eq. (\ref{eqEVC}), though our final  results are more general. 
The main tool is Eq. 
(\ref{eqG09})
 for $\langle \hat{{\cal S}}(p) \rangle$ which is evaluated
in the limit of small $p$, then transformed to $t$, a standard
Tauberian  procedure valid
in the long time limit. From here, as before, we get $\langle \overline{E}_k(t) \rangle \sim \langle {\cal S}(t) \rangle/t$.  

Eq. (\ref{eqG09}) is split into two terms
\begin{equation}
\langle {\cal S}(p) \rangle = \hat{H}_1 (p) + \hat{H}_2 (p)
\label{eqNI06}
\end{equation} 
with
\begin{equation}
\hat{H}_1 (p)  = - {{\partial \hat{\phi_2} (u,p) \over \partial u} |_{u=0}\hat{\Phi}_2 (0, p) 
\over \left[ 1 - \hat{\phi}_2 (0,p) \right]^2 },
\ \ \ \ \ \ \
\hat{H}_2(p) = 
- { {\partial \hat{\Phi}_2 \left(u,p\right) \over \partial u} |_{u=0} \over 1 -\hat{\phi}_2 (0,p)}.
\label{eqNI07}
\end{equation}
When $1/3<\gamma$ the derivative $\partial_u \hat{\phi}_2 (u,p)|_{u=0}$ is the mean action increment $\langle s \rangle$  when $p\to 0$. However,  here 
it diverges, since we consider $0<\gamma<1/3$. Recall that $\hat{\Phi}_2(u,p)$ is associated with the probability of not jolting namely the term $\hat{H}_2(p)$ is stemming from contribution to the total action from the backward recurrence time.
In the limit, it was negligible for $1/3< \gamma$ but here both terms are important. 

 To start we rewrite Eq. 
(\ref{eqFLUC02})
\begin{equation}
\hat{\phi}_2 (u, p ) = 1-  {1 \over v_{{\rm max}}} \int_0 ^{v_{{\rm max}}} \left( 1 -  {R(v) \over p + R(v) + u v^2} \right) {\rm d} v = 1 - {1 \over v_{{\rm max}}} \int_{0} ^{v_{{\rm max}}}
{ p + u v^2 \over p + R(v) + u v^2} {\rm d} v. 
\label{eqNI08}
\end{equation}
We set $u$ to zero and change variables according to $v^\alpha = (p c)^{1/\alpha} z$ and then get
\begin{equation}
\hat{\phi}_2 (0, p ) = 1-  { (p c)^{1/\alpha}  \over v_{{\rm max}}} \int_0 ^{v_{{\rm max}}/(p c)^{1/\alpha}} 
{ {\rm d} z \over 1 + z^\alpha} .
\label{eqNI09}
\end{equation}
The upper limit is then taken to be infinity and we find in the small
$p$ limit $\hat{\phi}_2(0, p) \sim 1 - ( p c)^\gamma \pi \gamma / v_{{\rm max}} \sin \pi \gamma$. The first term here is the normalisation, the second indicates that the
mean trapping time is diverging, more precisely this well known Tauberian
relation 
comes from the  fat tail of the waiting
time PDF 
Eq. (\ref{ctrwaa}).
In these calculations and those which now follow
 we use three integrals
\begin{equation}
\int_{0} ^{\infty} {{\rm d} z \over 1 + z^{\alpha} } 
= {(\pi/\alpha) \over \sin (\pi/\alpha) } ,  \  \ \ 
\int_0 ^{\infty} {\alpha z^{\alpha +2} {\rm d} z \over (1 + z^{\alpha})^2 } =
{(3 \pi/\alpha)   \over  \sin (3 \pi /\alpha)}, \ \mbox{and} \ \ \
\int_0 ^\infty {\alpha^2 z^2 {\rm d} z \over (1 + z^\alpha)^2 } = { \pi(\alpha-3) \over  \sin (3 \pi /\alpha) }, 
\label{eqNI10}
\end{equation} 
all valid in the regime of interest. 

Using the same tricks we find 
\begin{equation} 
- \partial_u \hat{\phi}_2 (u,p) |_{u=0}
\sim \left( { c \over v_{{\rm max}}} \right) \left(  p c \right)^{3 \gamma-1} \int_0 ^\infty { z^{\alpha+2 } {\rm d} z \over (1 + z^\alpha)^2 } , 
\label{eqNI11}
\end{equation} 
and 
\begin{equation} 
- \partial_u \hat{\Phi}_2 (u,p) |_{u=0} \sim { c^2 \over v_{{\rm max}}} \left( p c \right)^{3 \gamma-2} \int_0 ^\infty {z^2 {\rm d} z \over (1 + z^\alpha)^2  }.
\label{eqNI12}
\end{equation} 
Inserting Eqs. 
(\ref{eqNI11},
\ref{eqNI12})
in Eq. 
(\ref{eqNI07})
we find the small $p$ limit of $\hat{H}_1(p)$ and $\hat{H}_2(p)$. 
It is now easy to transform from $p$ to time $t$ and find
\begin{equation}
\langle \overline{E}_k \rangle = { \langle {\cal S} (t) \rangle \over t} \sim
\left( { c \over t} \right)^{2 \gamma} \left[
\underbrace{ { 3 \gamma \sin(\pi \gamma) \over \Gamma(2 - 2 \gamma) \sin (3 \pi \gamma) } }_{H_1(t)} + 
\underbrace{ { \sin (\pi \gamma) (1 - 3 \gamma) \over \Gamma(2 - 2 \gamma) \sin (3 \pi \gamma) }}_{H_2(t) } \right]. 
\label{eqNI12}
\end{equation} 
The first term vanishes in the limit of $\gamma\to 0$, namely in that limit
the time interval $(0,t)$ is effectively collision free, in such a way that the dominating
contribution is the backward recurrence time \cite{Longest,Marc} 
(the $H_2(t)$ term). 
This means roughly speaking, 
that in this limit the backward recurrence time is equal to the 
measurement time. 
In contrast when $\gamma\to 1/3$ marking the transition to the integrability
of the energy, namely the Darling-Kac 
phase, the first term which stems from many jumps is diverging. From
Eq. (\ref{eqNI12}) we get Eq. 
(\ref{TEA E_k}).

\subsection{EB parameter}

To obtain the $\mbox{EB}$ parameter we  use Eq. (\ref{eqEBp01}) to find the variance of ${\cal S}$ in the long
time limit. We consider the  four terms $I_j(p)$ with $j=1,..,4$  defined in Eq. (\ref{eqEBp01})  
in the limit $p \to 0$. 
The calculations are some what similar to those in the previous subsection though now
we need to consider second order derivatives with respect to $u$ for example
we find
\begin{equation}
\hat{\Phi}_2''(0,p)\sim {2 (p c)^{5/\alpha}  \over v_{{\rm max}} p^3}  \int_0 ^\infty { z^4 {\rm d} z \over (1 + z^{\alpha})^3}.
\label{eqEBp02}
\end{equation} 
and using  Eq. (\ref{eqNI09})
\begin{equation}
I_1(p) \sim { 2 c^3 (p c)^{4 \gamma-3} \int_0 ^\infty { z^4 \over (1 + z^\alpha)^3 } {\rm d} z \over (\pi \gamma)/\sin (\pi \gamma)}.
\label{eqEBp03}
\end{equation}
Notice that $I_1(p)$ is independent of $v_{{\rm max}}$ in this limit, and so are the remaining terms $I_2(p),I_3(p)$ and $I_4(p)$ 
which are given by
\begin{equation}
I_2(p) \sim 2 c^3 (p c)^{4 \gamma-3} 
{\int_0 ^\infty { z^2 \over (1 + z^\alpha)^2} {\rm d} z \int_0 ^\infty { z^{\alpha+2} \over (1 + z^\alpha)^2 } {\rm d} z \over  [(\pi \gamma) / \sin(\pi \gamma)]^2},
\label{eqEBp04}
\end{equation}
\begin{equation}
I_3(p) \sim 2 c^3 ( p c)^{4 \gamma -3} { \int_0 ^\infty {z^{\alpha +4} \over (1 + z^\alpha)^3} {\rm d} z \over \pi \gamma/\sin \pi \gamma} ,
\label{eqEBp05}
\end{equation}
and similarly 
\begin{equation}
I_4(p) \sim 2 c^3 (p c)^{4 \gamma-3} { \left[  \int_0 ^\infty { z^{\alpha +2} \over (1 + z^\alpha)^2} {\rm d} z   \right]^2  \over \left[ \pi \gamma/ \sin \pi \gamma\right]^2}.
\label{eqEBp06}
\end{equation}
Inverting to the time domain we see that $\langle S^2 (t) \rangle \propto t^{ 2-4 \gamma}$.
The integrals
in Eqs. 
(\ref{eqEBp03}
-\ref{eqEBp06}) are tabulated in Mathematica, so
summing  all the  four  terms 
in Eq. (\ref{eqEBp01})
and using $\langle S(t) \rangle$, 
Eq. 
(\ref{eqNI12}),
we obtain the variance $\langle S^2(t) \rangle - \langle S(t)\rangle^2$ 
and this 
after normalization
 yields the $\mbox{EB}$ parameter
Eq. 
(\ref{eqNI13}).

\section{Distribution of time averages in the non-integrable phase}
\label{Sec7}

We will now obtain the PDF of the random variable
$\Upsilon= {\cal S}(t) / \langle {\cal S}(t) \rangle =
 {\overline E}_k(t) /\langle \overline{E}_k (t) \rangle$, namely the normalised time averaged kinetic energy, in the phase when this observable is non-integrable with respect to the infinite density. Recall, that
when the energy is integrable, we obtain the universal Mittag-Laffler law 
Eq. (\ref{eqML}). Unlike the latter case, the PDF of $\Upsilon$ denoted 
$P_\alpha(\Upsilon)$ will now depend on the microscopical details of the model, in particular the PDF of the speed  after collisions  $f(v)$.
 Here we find $P_\alpha (\Upsilon)$ for the model under study, namely the case where $f(v)$ is uniform. The analysis does not allow us to obtain a
general solution for $P_\alpha (\Upsilon)$ 
for all $3<\alpha$ and we mainly revert to approximations. 
This is unlike the variance, given in terms of the EB parameter,
Eq.
(\ref{eqNI13}) which was calculated exactly. 
Since the calculations are lengthy, here we provide the outline of the theory, 
focusing on three cases, $\alpha=4$, $\alpha=6$ and $\alpha\to \infty$.
Comparing the semi-analytical solution to simulations we gain insight
on a new type of transition, which shows up as a sudden blow up
of $P_\alpha(\Upsilon)$ for $\Upsilon\to 0$.  The effect is certainly
not found for the Mittag-Leffler distribution, within the integrable phase $\alpha<3$. 

The double Laplace transform of the action propagator $P(S,t)$ is given by $%
\widehat{P}(u,p)$ and the Montroll-Weiss type  equation
Eq. (\ref{eqG07}).
The technical problem is to invert this solution in the limit of long
times corresponding to the Laplace variable  $p\to 0$ being small.
The functions 
$P({\cal S},t)$ and  $%
\widehat{P}(u,p)$  attain scaling forms, denoted
$P({\cal S},t) \sim t^{1-\alpha/2} f_\alpha ( {\cal S} / t^{1-2/\alpha})$ and
$P(u,p) \sim (1/p) g_\alpha ( u/p^{1- 2/\alpha})$. Here the limit under  study
is $t \to \infty$ and ${\cal S} \to \infty$ the ratio 
${\cal S} / t^{1-2/\alpha}$ remaining fixed and similarly in Laplace space.  
Note that we showed in Eq. 
(\ref{eqNI12}) that $\langle {\cal S}(t) \rangle \propto t^{1-2/\alpha}$, hence
the scaling of ${\cal S}$ with time  we use here is consistent with that observation. The two scaling solutions are related by the laws of Laplace transform
\begin{equation}
\frac{1}{p}g_{\alpha }(\frac{u}{p^{1-2/\alpha }})=\int\limits_{0}^{\infty
}d {\cal S}\int\limits_{0}^{\infty }dt\;e^{-u{\cal S}}e^{-pt}\frac{1}{t^{1-2/\alpha }}%
f_{\alpha }(\frac{{\cal S}}{t^{1-2/\alpha }}).  \label{double Laplace scaling}
\end{equation}
Substituting $x=\frac{{\cal S}}{t^{1-2/\alpha }}$ and setting $p=1$ turns Eq. (\ref%
{double Laplace scaling}) into the integral equation%
\begin{equation}
g_{\alpha }(y)=\int\limits_{0}^{\infty }dx\;K_{\alpha }(yx)f_{\alpha }(x)
\label{Fredholm}
\end{equation}%
relating $g_{\alpha }(y)$ and $f_{\alpha }(x)$ via this convolution
transform with kernel
\begin{equation}
K_{\alpha }(x)=\int\limits_{0}^{\infty }dt\;e^{-xt^{1-2/\alpha }}e^{-t}.
\label{kernel}
\end{equation}%
For the scaling function $g_{\alpha }(y)$, with $\alpha >3$, i.e. in the
non-Darling-Kac phase, we obtain from the  $p\rightarrow 0$ limit of 
$\widehat{P}(u,p)$  
Eq. (\ref{eqG07})
 the following exact form
\begin{equation}
g_{\alpha }(y)=\frac{\int\limits_{0}^{\infty }\frac{1}{1+yz^{2}+z^{\alpha }}%
dz}{\int\limits_{0}^{\infty }\frac{1+yz^{2}}{1+yz^{2}+z^{\alpha }}dz}.
\label{gy exact}
\end{equation}
The goal is to obtain from this by inversion of Eq. (\ref{Fredholm}) the
scaling function $f_{\alpha }(x)$, which is a rescaled version of the
limiting probability density $P_{\alpha }(\Upsilon )$ of the normalized time
average $\Upsilon $. Such an inversion can be achieved in
principle by a Mellin transform of both sides of Eq. (\ref{Fredholm})
resulting in [\cite{Polyanin}, p.997]
\begin{equation}
\widetilde{g}_{\alpha }(s)=\widetilde{K_{\alpha }}(s)\widetilde{f}_{\alpha
}(1-s),  \label{Mellin product}
\end{equation}
where $\widetilde{g}_{\alpha }(s)=M[g_{\alpha
}(y);s]=\int\limits_{0}^{\infty }dy\;g_{\alpha }(y)y^{s-1}$ is the Mellin
transform of $g_{\alpha }(y)$, and $\widetilde{K_{\alpha }}(s)$ and $%
\widetilde{f}_{\alpha }(s)$ is defined analogously. Solving Eq. (\ref{Mellin
product}) for $\widetilde{f}_{\alpha }(s)$ and applying the inverse Mellin
transformation gives
\begin{equation}
f_{\alpha }(x)=M^{-1}[\widetilde{f}_{\alpha }(s);x]=M^{-1}[\frac{\widetilde{g%
}_{\alpha }(1-s)}{\widetilde{K}_{\alpha }(1-s)};x],  \label{inverse Mellin}
\end{equation}%
and finally by rescaling the desired limit density is obtained as%
\begin{equation}
P_{\alpha }(\Upsilon)=C_{\alpha }f_{\alpha }(x)|_{x=C_\alpha \Upsilon},  \label{rhofx}
\end{equation}%
where $C_{\alpha }=\left\langle x\right\rangle _{f_{\alpha
}}=\int\limits_{0}^{\infty }dx\;xf_{\alpha }(x)$. The rescaling implies
that the mean is $\left\langle \Upsilon \right\rangle _{P_{\alpha
}}=\int\limits_{0}^{\infty }d \Upsilon\;\Upsilon P_{\alpha }(\Upsilon)=1$, as requested. While in
principle along these steps the problem of calculating the limit
distribution $P_{\alpha }(\Upsilon)$ is solved, a fully analytic solution is
available only in two cases, $\alpha =4$ and $\alpha =\infty $. For $\alpha
=4$ the integrals in Eq. (\ref{gy exact}) can be evaluated by residue
calculus yielding after some calculations the simple result
\begin{equation}
g_{4}(y)=\frac{1}{1+y},  \label{g4 final}
\end{equation}%
with Mellin transform $\widetilde{g}_{4}(s)=\Gamma (s)\Gamma (1-s)$. Since
the Mellin transform $\widetilde{K_{\alpha }}(s)$ of the integral kernel is
also known in full generality, $\widetilde{K}_{\alpha }(s)=\Gamma (s)\Gamma
(1-(1-2/\alpha )s)$, we get the quotient in Eq. (\ref{inverse Mellin}), and
in addition we can invert from Mellin space to obtain $f_{4}(x)=\frac{1}{%
\sqrt{\pi }}e^{-\frac{x^{2}}{4}}$. The scaling factor $C_{4}=\left\langle
x\right\rangle _{f_{4}}$ follows from the general relation between the $n$%
-th derivative $g_{\alpha }^{(n)}(y=0)$ and the moments $\left\langle
x^{n}\right\rangle _{f_{\alpha }}$ of $f_{\alpha }(x)$
\begin{equation}
g_{\alpha }^{(n)}(0)=K_{\alpha }^{(n)}(0)\left\langle x^{n}\right\rangle
_{f_{\alpha }}=(-1)^{n}\Gamma (1+n(1-\frac{2}{\alpha }))\left\langle
x^{n}\right\rangle _{f_{\alpha }},  \label{moments general}
\end{equation}%
which follows directly from Eq. (\ref{Fredholm}). For $\alpha =4$ we get $%
C_{4}=\left\langle x\right\rangle _{f_{4}}=\frac{2}{\sqrt{\pi }}$, so that
we obtain for $P_{4}(\Upsilon)$ according to Eq. (\ref{rhofx}) a half Gaussian
distribution
\begin{equation}
P_{4}(\Upsilon)=\frac{2}{\pi }e^{-\frac{\Upsilon^{2}}{\pi }}  \label{rho4}
\end{equation}
as an exact result.
It is merely a coincidence, that 
 this half Gaussian is found also in the  Mittag-Leffler phase,
when $\alpha=2$, see Fig. 
\ref{fig5}. 
We can proceed similarly for $\alpha \rightarrow
\infty $, because all integrals and Mellin transforms are known exactly also
in this case yielding
\begin{equation}
g_{\infty }(y)=\frac{1}{\sqrt{y}}\arctan \sqrt{y}  \label{gyinf}
\end{equation}%
and eventually 
\begin{equation}
P_{\infty }(\Upsilon)=\frac{1}{2\sqrt{3}}\frac{1}{\sqrt{\Upsilon}}\theta (3-\Upsilon),
\label{rhoinf}
\end{equation}
where we use the step function. 
This result diverges at the origin $\Upsilon\to 0$, and we will soon show
that this is valid for any $\alpha>4$.
Note that PDF is cutoff sharply at $\Upsilon=3$ an effect which
is related to the underlying uniform distribution of $v$, Eq.
(\ref{eqFV}).
More precisely, we may explain this result, by noting that
the atom maintains a constant velocity for practically all the duration
of the experiment, since $\alpha\to \infty$ or $\gamma=1/\alpha \to 0$.
Then $\Upsilon=v^2/\langle v^2 \rangle$ and using the uniform PDF
of velocities Eq. (\ref{eqFV}) we get Eq. (\ref{rhoinf}). 
In Fig. \ref{a50} we present simulation results  for $\alpha=50$
and compare them to theory $\alpha
\to \infty$.  These match well
with the limiting PDF $P_\infty (\Upsilon)$, the exception is
that the histogram
is smeared and does not show  the step like structure of
limiting PDF which is found at $\Upsilon=3$.

\begin{figure}\begin{center}
\epsfig{figure=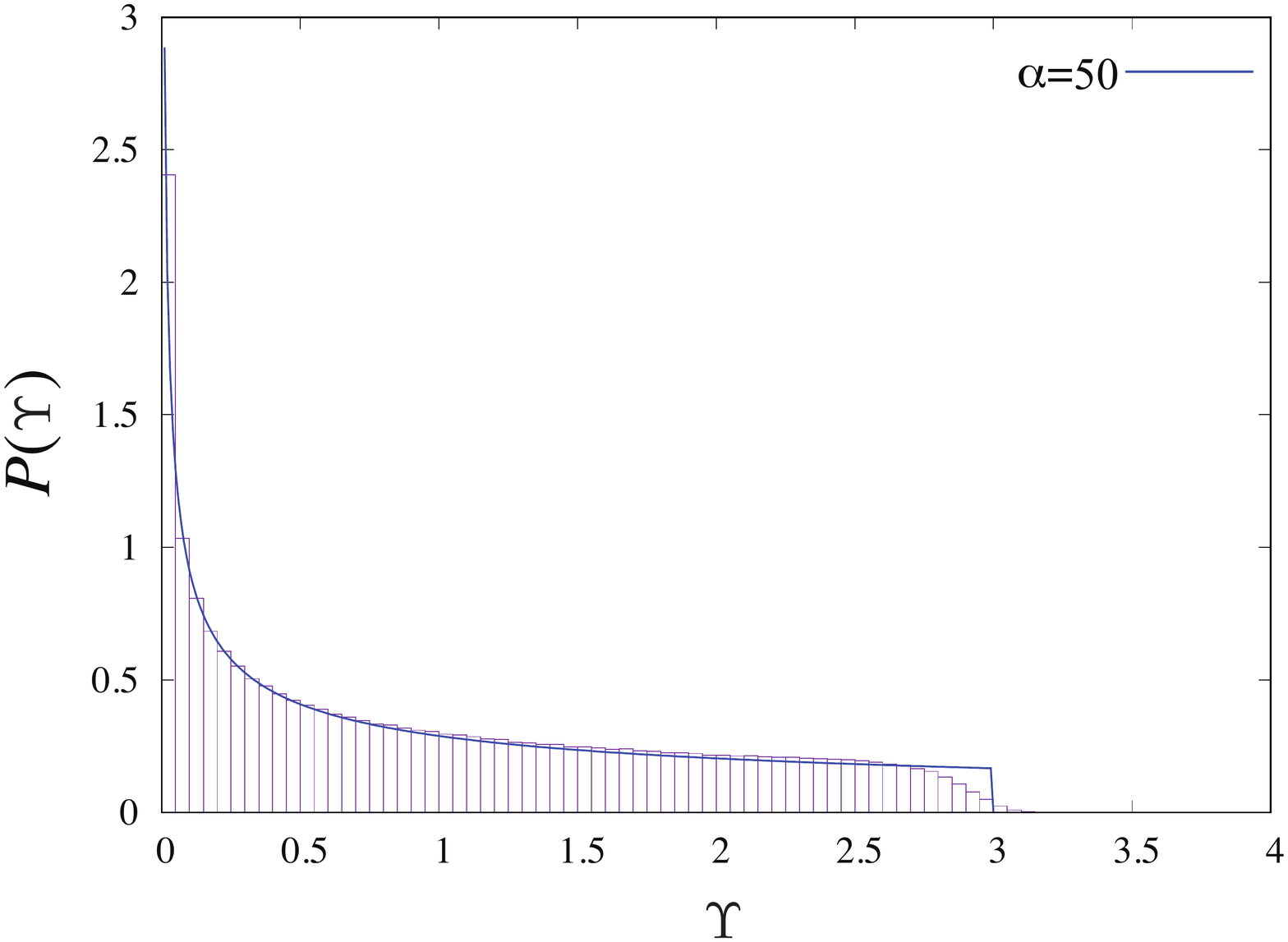, width=0.51\textwidth,trim=0mm 0mm 0mm 0mm, clip}
\end{center}
\caption{
Histogram of the normalized time averaged energy $P_\alpha(\Upsilon)$, obtained from
numerical simulations with $\alpha=50$ is compared with the limiting PDF
Eq. 
(\ref{rhoinf}). In the limit the PDF is sharply 
 cutoff at $\Upsilon=3$ an effect which is smeared out
with the finite time, finite $\alpha$ simulations. 
}
\label{a50}
\end{figure}

For the experimentally also relevant case $\alpha =6$ we can still get from
Eq. (\ref{gy exact}) by residue calculus a fully analytic expression for $%
g_{6}(y)$, but the result is very lengthy involving 3rd roots etc., and it
does not simplify as in the case  $\alpha =4$. Therefore one cannot
analytically calculate its Mellin transform. This led us to find an
approximate scaling function $g_{6}^{\approx }(y)$, which deviates only
little from the true function $g_{6}(y)$, but for which the Mellin transform
is known analytically. The general ideas is to match in $g_{6}^{\approx }(y)$
the true small $y$-behavior and the large $y$-asymptotics of $g_{6}(y)$,
which we know analytically from an analysis of Eq. (\ref{gy exact}). The
simple form 
\begin{equation}
g_{6}^{\approx }(y)=(1+\frac{2}{3}y)^{-\frac{3}{4}}  \label{g6yapprox}
\end{equation}%
shares with the exact function $g_{6}(y)$ the identical first and second
derivative at $y=0$, and the exponent of the asymptotic behavior for $%
y\rightarrow \infty $. The relative deviation of  $g_{6}^{\approx }(y)$ from
$g_{6}(y)$ is strictly less than 4.2\%. This function $g_{6}^{\approx }(y)$
can be Mellin transformed  \cite{Prudnikov}, and leads via  Eq. (\ref{inverse
Mellin}) and subsequent rescaling to an analytical expression for $%
P_{6}^{\approx }(\Upsilon)$, which can be expressed in terms of a Fox-H function
\cite{Fox}
as
\begin{equation}
P_{6}^{\approx }(\Upsilon)=\frac{3}{4\Gamma (\frac{5}{3})\Gamma (\frac{3}{4})}%
H_{1,1}^{1,0}\left( \frac{3}{4\Gamma (\frac{5}{3})}\Upsilon\left\vert
\begin{array}{c}
(\frac{1}{3},\frac{2}{3}) \\
(-\frac{1}{4},1)%
\end{array}%
\right. \right) .  \label{rho6xapproxfinal1}
\end{equation}%
As mentioned, after rewriting this expression in terms of a Meijer G-function
 [\cite{Prudnikov},
p.629], we can plot the result with Mathematica, see Fig. \ref{a6}.

\begin{figure}\begin{center}
\epsfig{figure=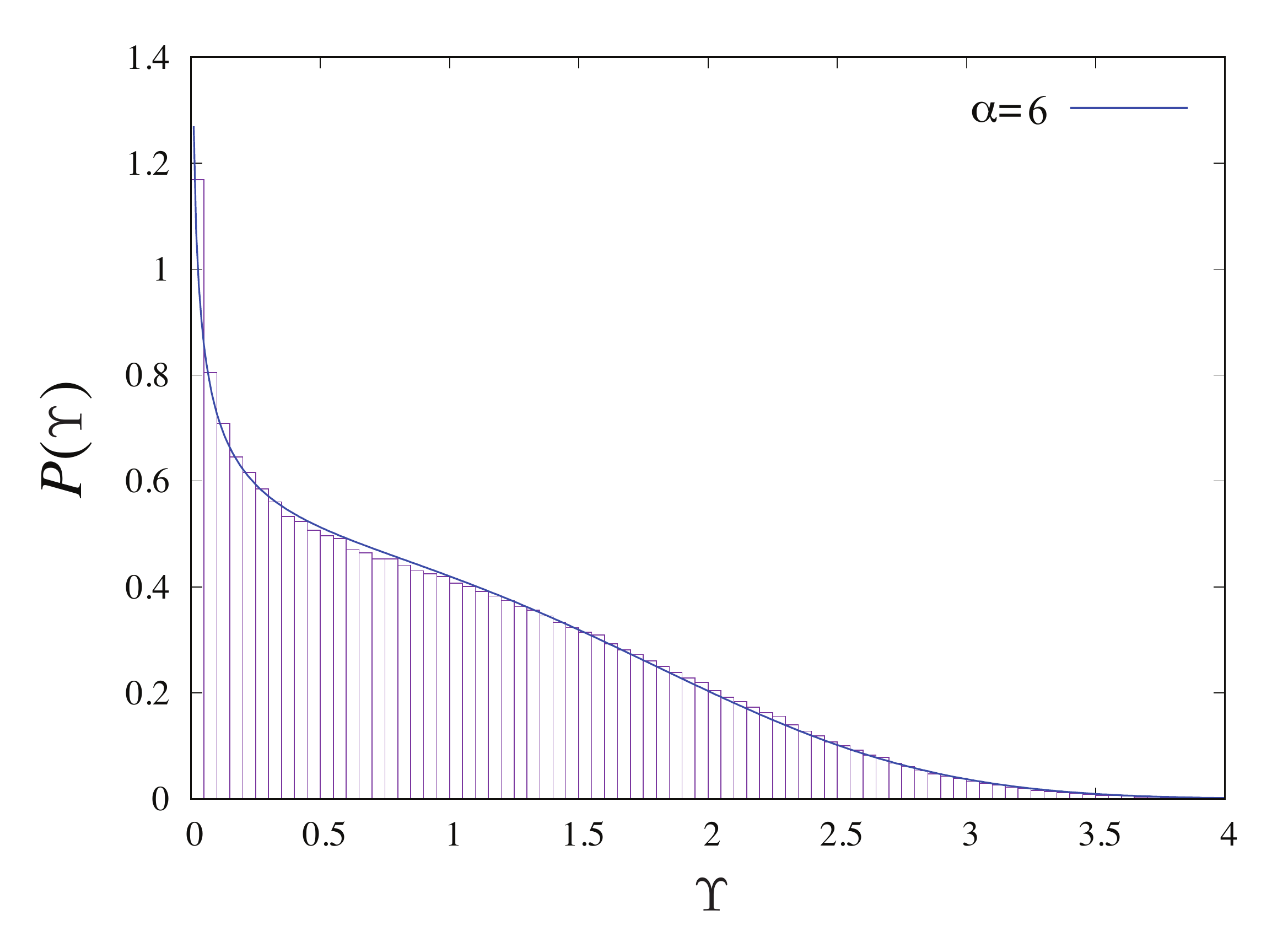, width=0.51\textwidth,trim=0mm 0mm 0mm 0mm, clip}
\end{center}
\caption{
Simulations and theory Eq. (\ref{rho6xapproxfinal1}) for the distribution of the
time averaged energy  $P_\alpha(\Upsilon)$, 
for  $\alpha=6$.
Here the observable is non-integrable with respect to the infinite density, and
$P_6(\Upsilon)$ diverges at $\Upsilon\to 0$, indicating very long sticking
times close to zero speed, 
 for some of the atoms which have very low energy compared to the
mean. 
}
\label{a6}
\end{figure}

\subsection{Accumulation effect for $\alpha>4$}

 We mentioned already that $P_\alpha (\Upsilon)$ diverges at the origin
$\Upsilon \to 0$ for any $\alpha>4$ and explained
that  this means that a large population
of particles remain slow for the whole duration of the experiment.
Now we wish to characterise this effect more precisely.

It is easy to see that the exact asymptotic behavior of $g_{\alpha }(y)$ for $y\rightarrow \infty $
is given by%
\begin{equation}
g_{\alpha }(y)\sim ry^{-\eta },  \label{asygy}
\end{equation}%
with decay exponent 
\begin{equation}
\eta =\eta (\alpha )=\frac{\alpha }{2(\alpha -2)}  \label{decayexp}
\end{equation}%
and decay amplitude 
\begin{equation}
r=r(\alpha )=\frac{\alpha -2}{2}\sin \frac{\pi }{\alpha -2}.
\label{decayamp}
\end{equation}
From that the small $x$-asymptotics of $f_{\alpha }(x)$ follows exactly as%
\begin{equation}
f_{\alpha }(x)\sim t(\alpha )x^{-s_{0}(\alpha )}  \label{asyfx}
\end{equation}%
with exponent
\begin{equation}
s_{0}(\alpha )=1-\eta (\alpha )=1-\frac{\text{ }\alpha }{2(\alpha -2)}=\frac{%
\alpha -4}{2\alpha -4},  \label{s0}
\end{equation}%
and amplitude  
\begin{equation}
t(\alpha )=\frac{r(\alpha )}{\sqrt{\pi }\Gamma (\eta (\alpha ))}=\frac{\frac{%
\alpha -2}{2}\sin \frac{\pi }{\alpha -2}}{\sqrt{\pi }\Gamma (\frac{\alpha }{%
2(\alpha -2)})} . \label{famp}
\end{equation}%
The following table gives an overview for the resulting behavior of $%
f_{\alpha }(x\rightarrow 0)\sim $ $t(\alpha )x^{-s_{0}(\alpha )}$ and of the
limit distribution $\rho _{\alpha }^{\ast }(x\rightarrow 0)=C_{\alpha
}f_{\alpha }(C_{\alpha }x\rightarrow 0)$. With the explicit form of $%
C_{\alpha }$ derivable from the exact form of $g_{\alpha }^{\prime }(0)$  
\begin{equation}
C_{\alpha }=\left\langle x\right\rangle _{f_{\alpha }}=\frac{\sin (\pi
/\alpha )}{\sin (3\pi /\alpha )}\frac{1}{\Gamma (2-\frac{2}{\alpha })}=\frac{%
1}{1+2\cos \frac{2\pi }{\alpha }}\frac{1}{\Gamma (2-\frac{2}{\alpha })}
\label{C_alpha}
\end{equation}
this yields
\begin{equation}
\begin{array}{ccc}
\alpha  & f_{\alpha }(x\rightarrow 0) & P_{\alpha }(x\rightarrow
0) \\ 
\alpha \rightarrow \infty  & \frac{1}{2}x^{-\frac{1}{2}} & \frac{1}{2\sqrt{3}%
}x^{-\frac{1}{2}} \\ 
\alpha =50 & \frac{24\sin (\frac{\pi }{48})}{\sqrt{\pi }\Gamma (\frac{25}{48}%
)}x^{-\frac{23}{48}} & \frac{24\sin (\frac{\pi }{48})}{\sqrt{\pi }\Gamma (%
\frac{25}{48})[(1+2\cos (\frac{\pi }{25}))\Gamma (\frac{49}{25})]^{\frac{25}{%
48}}}x^{-\frac{23}{48}} \\ 
\alpha =6 & \frac{\sqrt{2}}{\sqrt{\pi }\Gamma (\frac{3}{4})}x^{-\frac{1}{4}}
& \frac{1}{2^{\frac{1}{4}}\sqrt{\pi }\Gamma (\frac{3}{4})\Gamma (\frac{5}{3}%
)^{\frac{3}{4}}}x^{-\frac{1}{4}} \\ 
\alpha =5 & \frac{3\sqrt{3}}{4\sqrt{\pi }\Gamma (\frac{5}{6})}x^{-\frac{1}{6}%
} & \frac{3\sqrt{3}[\frac{1}{2}(3-\sqrt{5})]^{\frac{5}{12}}}{4\sqrt{\pi }%
\Gamma (\frac{5}{6})\Gamma (\frac{8}{5})^{\frac{5}{6}}}x^{-\frac{1}{6}} \\ 
\alpha =4 & \frac{1}{\sqrt{\pi }}x^{0} & \frac{2}{\pi }x^{0} \\ 
\alpha =\frac{7}{2} & \frac{3\sqrt{3}}{8\sqrt{\pi }\Gamma (\frac{7}{6})}x^{%
\frac{1}{6}} & \frac{3\sqrt{3}}{8\sqrt{\pi }\Gamma (\frac{7}{6})}\left[ 
\frac{\cos (\frac{3\pi }{14})}{\sin (\frac{\pi }{7})\Gamma (\frac{10}{7})}%
\right] ^{\frac{7}{6}}x^{\frac{1}{6}} \\ 
\alpha =\frac{10}{3} & \frac{\sqrt{2}}{3\sqrt{\pi }\Gamma (\frac{5}{4})}x^{%
\frac{1}{4}} & \frac{(1+\sqrt{5})^{\frac{5}{2}}}{12\sqrt{\pi }\Gamma (\frac{5%
}{4})\Gamma (\frac{7}{5})^{\frac{5}{4}}}x^{\frac{1}{4}} \\ 
\begin{array}{c}
\alpha =3+\varepsilon , \\ 
\varepsilon \rightarrow 0%
\end{array}
& \varepsilon x^{\frac{1}{2}-\varepsilon } & (\frac{2\sqrt{3}}{\pi \Gamma (%
\frac{7}{5})})^{\frac{3}{2}}\varepsilon ^{-\frac{1}{2}}x^{\frac{1}{2}%
-\varepsilon }.
\end{array}
\label{table}
\end{equation}
This array of equations shows that $\alpha =4$ takes a special role as it separates
diverging behavior of $f_{\alpha }(x)$ near $x=0$ for $\alpha >4$ from
vanishing behavior for $3<\alpha <4$.  Note also the diverging amplitude in $%
\rho _{3+\varepsilon }^{\ast }(x)$ as $\varepsilon \rightarrow 0$ in
contrast to the vanishing amplitude of $f_{3+\varepsilon }(x)$, which is a
consequence of the diverging scaling factor $C_{3+\varepsilon }\sim
\varepsilon ^{-1}$.

\subsection{Physical consequences}

We discovered for $\alpha>4$, an accumulation effect,
namely the divergence of the PDF of the time averages,
found at low energies
e.g. where $\alpha=6$ and $\Upsilon\to 0$.
This means that a significant population of atoms
remains at small speeds for the whole duration of the experiment.
In turn, this is useful when one wishes to reduce scattering or spatial spreading, namely holding
atoms close to the dark zero momentum state for long durations.
 Thus,
while for the optimization of the commonly used
 relaxation time of the full width at half maximum (FWHM) of the velocity packet,
which decays as $t^{-1/\alpha}$,
one should consider small values of $\alpha$ to obtain fast relaxation
 (say $\alpha=2$),
to maintain some of the population with small
kinetic energy for long durations,
large values of $\alpha$ (say $\alpha=6)$ are beneficial, as the trapping times
become statistically longer.
Surprisingly,  $\alpha=4$ marks a quantitative transition
of the low energy
statistics, which we discovered from the analysis of the time
averages.

\section{Perspective}

 The rate of escape from the velocity trap, $R (v) \propto v^{\alpha}$ for
small $v$, implies that for  laser cooled systems the  mean escape
time  is infinite when $1<\alpha$ or equivalently $\gamma<1$. 
From the point of view of cooling this is 
an advantage, in the sense that typical speeds are low (nano-Kelvin
regime). At the same time this leads to the applicability 
of infinite ergodic theory, including the non-normalisable measure.
The system shares some features which are  similar to glassy dynamics,
in particular the trap model \cite{WEB}.
 In that model, we have a density of states
$\rho(E)= \exp[- E/T_g]/T_g$ where $E>0$ are the trap depths
and $T_g$ a measure of disorder. The  system is in contact
with a heat bath at temperature $T$. We will not go into the details
of the anomalous dynamics in this model, however we point out that also
here we encounter a non-normalisable state. The partition function
is $Z=\int_0 ^\infty \rho(E) \exp( E/T) {\rm d} E$ (here $k_B=1)$ and hence it
diverges when $T_g>T$.  This low temperature glassy phase 
 also corresponds to the  case where the mean trapping time
diverges and where one finds anomalous kinetics. 
Thus, for both the sub-recoiled system and  the trap model,
we find diverging mean trapping times and also the blow up of the normalisation
of the usual steady state. Bouchaud described such systems as exhibiting
weak ergodicity breaking, since one has exploration of phase space although time
and ensemble averages differ \cite{WEB}. Mathematicians, use the term infinite
ergodic theory, since they realise that the non-normalisable measure is
the key ingredient of the theory. 
Further, this non-normalisable state  is related to the normalised
distribution $\rho(v,t)$  see Eq. 
(\ref{eq12}),
and it is approached from a broad class
of initial conditions. Thus, some actually  call the dynamics ergodic, 
i.e. the term infinite ergodic theory implies the dynamics is ergodic,
while in the physics literature others  describe it as non-ergodic.
 In short one should distinguish between the operational
definition of ergodicity, time and ensemble averages coincide, and
the fact that in the long-time limit a unique density is approached,
be it normalised or not. 
We conclude that
 weak ergodicity breaking and infinite ergodic theory are deeply related.
The statistical theory applies both, to models in a non-equilibrium 
setting
like laser cooled atoms, but also to systems with  a  canonical
 Boltzmann-Gibbs measure  
even if the latter is not normalised \cite{Erez}.

What are the consequences for laser cooling?
 Remarkably,  using Eq.
(\ref{eqNI01})
we conclude that the most efficient cooling, in the sense of the fastest relaxation of the mean energy, is found for $\gamma=1/3$.
Thus the transition in the ergodic properties of the system
 investigated here, which takes place when $\alpha=3$ or $\gamma=1/3$,
 is physically connected to the optimal
cooling of energy.
This is not a coincidence, namely at the transition point the time dependence changes, though the fact that this point is optimal seems to us as merely
good luck. 
In contrast, for the FWHM
of the velocity packet \cite{Reichel}, we do not have such an optimum.
Instead, as mentioned, it
 decays like $t^{-1/\alpha}$ favouring small
values of $\alpha$ for faster relaxation \cite{Reichel}.
Thus the classification of
an observable as either  integrable (energy, $\gamma>1/3$) or non-integrable
(energy $\gamma<1/3$, FWHM) with respect to the infinite invariant measure
 is crucial, both mathematically and
physically.
We should note that the FWHM is not a dynamical observable, in the sense that
it cannot be obtained as a functional of a single particle  path.

 In the context of sub-recoil laser cooling our work raised a few questions
and here we point out possible extensions. 
\begin{itemize}
\item[1.]
It is a challenge to see if quantum Monte Carlo simulations \cite{CT}
 can be used
to investigate numerically the non-normalisable state and the time and ensemble
averages. 
\item[2.] Experimentally finding the infinite density is  another obvious challenge. Single atom experiments yield direct insights
on the trajectories and the time averages \cite{Katori,Stroescu,Hohmann,Widera}. 
\item[3.] In our examples we used simple forms for $f(v)$ and $R(v)$.
It is important to realise that our main results are generally valid,
like the applicability of infinite ergodic theory, though it is clear that
details do depend on the microscopical behaviour of these functions.
In this context we have recently considered \cite{preparation}
 other models, including the
case where the process is not renewed after each jolt. The main results
are left unchanged. 
\item[4.] What happens to such a gas of atoms in the presence of some
binding field, e.g. a harmonic trap? What is the pressure of the gas?
What will happen when we add interactions? Will that drive the system
to a true thermal state?
\item[5.] When we considered 
 time averages, the measurement starts
when the process is initiated (lower limit of the time integral is zero).
 Instead one may prepare the system at
time $t=0$, then wait until a time $t_a$ and only then perform a measurement,
i.e. time average in the interval $(t_a,t_a+t)$. In this case
we expect that statistical properties of the system will depend on the ageing time $t_a$.
One can then wonder whether the infinite measure will play an important
role also under these conditions? In this regard we may be optimistic,
see the  modification
of Darling-Kac theorem to the ageing regime, in the context
of deterministic  dynamics \cite{Akimoto}.

\item[6.]  Sisyphus cooling is also described in terms of L\'evy processes
\cite{Castin,Zoller,RenzoniT,Sagi,Kessler1,Dechant,AghionPRX} 
and infinite {\em covariant} densities were studied  in this context
\cite{Kessler,Holz}. 
Hence the statistical framework of non-normalised states
is indeed widely applicable \cite{Renzoni}. 
However for Sisyphus cooling the physics is orthogonal
to the current one.  The Sisyphus friction force is vanishing for large $v$,
and thus the non-normalisable trait comes from the high speed particles
\cite{Kessler}.
Here, we have the opposite situation, the rate is anomalous for small
$v$. Technically this is related to the fact that for optical lattices
infinite covariant densities are studied, while here the focus on
infinite invariant densities. 

\item[7.] According to the model, the width of the velocity distribution
 shrinks with
time. And as mentioned in the text, this was indeed observed in experiments
till times of order milliseconds. Setting aside experiments, in the limit of 
$t \to \infty$ the model's predicts  a complete pile up 
at zero velocity, which seems to be a far fetched idea. 
We have thus considered an idealised situation, in fact, one could introduce
some cutoffs to the process  using $R(v)\propto v^{\alpha} + \epsilon$
where $\epsilon$ is very small. Such a cutoff could be important as
it would mean that in the very long time limit the system will eventually
relax to a normalised state.  Indeed, in the absence of cutoffs the time renewal process is scale free and hence it is a random fractal. 
Like any fractal in nature, cutoffs could
be important, as is well known.

\item[8.] As shown here clearly, and as well known more generally, the theory of infinite ergodic theory is a theory of observables. 
For example, in our study 
 the indicator function $I(v_a <v(t) <v_b)$ is integrable
with respect to the infinite density when $0<v_a$ and non-integrable 
if $v_a=0$. This is because of the non-integrable nature of the infinite
density at small $v$, which, as stressed, is related to the fact that $R(v)\sim v^\alpha$. The kinetic energy is integrable when $1/3 <\gamma=1/\alpha<1$,
 and this 
has consequences  for the ergodic properties of the process. One could consider
other observables like $|v|$, the main conclusions of this paper would be left
unchanged. 
It should be noted that also when the invariant measure is finite, and a usual
steady state exists an observable can be non-integrable,
e.g. in a thermal setting a particle in a harmonic trap, has a Boltzmann density 
proportional to $\exp( - k x^2 / 2 k_B T)$, hence an observable which might seem a bit weird
to some, like
${\cal O} (x) =\exp( x^4)$ is non-integrable.  
 The case of non-integrable observables with respect to the 
steady state, i.e. $\alpha<1$, 
 is of some theoretical  interest in the context of
the stochastic model under study. At first this might
seem academic since so far in experiment $\alpha>1$, however
 this could be of interest in dimensions
greater that unity, see below. An issue in our  mind is, whether  a specific
observable is {\em  physically interesting or measurable}, 
and we worked under the assumption
that energy is a physically worthy observable. 

\item[9.] We considered here the parent velocity
distribution $f(v)$ being a constant at small $v$, e.g. a
 uniform velocity distribution. In \cite{Bertin}
it was suggested that
$f(v) = \mbox{const}\ v^{d-1}$ for $v<v_{{\rm max}}$ and  $d$ is the dimension. 
Hence as mentioned  the focus of this paper was on one dimensional systems, mainly for the sake of simplicity \cite{Nir}. However, once again
the main conclusions of our paper are left unchanged, or more correctly,
 when minor adaptations are made, we may reach similar conclusions.
 For example, the infinite density in the general case,
 ${\cal I}_v(v) \propto v^{ - \alpha + d-1}$,
which is clearly non-normalisable when
$\alpha-d+1>1$. Hence the case $\alpha=2$
and $d=2$ is special, as it falls on the border between ergodicity in its
usual sense and infinite ergodic theory. Such cases are left for future work.

\end{itemize} 

\section{Summary}

Our starting point was the master equation for the speed of the particle
which was previously studied with several methods \cite{CT,Bertin}.
Here we highlighted the infinite measure ${\cal I}_v(v)$ which is a non-normalisable quasi-steady state of the system. To explore the ergodic properties of the
system we introduced a generalised
 L\'evy walk approach. 
This tool gives a Montroll-Weiss like  formula, Eq. 
(\ref{eqG07}),
which is a formal solution to the problem, but more importantly, it 
can be analysed in the long-time limit  giving
statistical information on the distribution of the action ${\cal S}(t)$ and
from it the distribution of the time average of the  energy $\overline{E_K}(t) = {\cal S}(t)/t$. 
Here we used the fact that between
collision events the momentum is conserved, so the speed is constant changing
abruptly at random times. With this method we are able to obtain the properties of the time averages which are functionals of the stochastic process. 
  We focused on the kinetic energy of the particles,
however, the approach presented here,
 is more general. Technically the increments
of the walk are action increments $s$, and the joint PDF of the increments
$s$ and the waiting times, are the basic ingredients of this coupled
walk. 

We find three phases
in the ergodic properties of the process. The case $\gamma>1$, corresponding to
a finite mean time between collisions,  was not considered here in detail,
since the standard ergodic theory  holds as the invariant measure is
normalisable. 
 In the regime $1/3<\gamma<1$ the Darling-Kac theorem holds,
for the observable of interest. This means that we may quantify the fluctuation
of the time averages using the Mittag-Leffler law, Eq.
(\ref{eqFLUC09}),
which is a
 universal type of statistical law in these type of problems.
 More precisely this theorem is valid for observables which are integrable with respect
to the infinite density.  Finally,
when $0<\gamma<1/3$, the energy observable
is non-integrable with respect to the infinite density. Here the mean action 
increment $\langle s \rangle$   diverges together
with the mean time between collisions. The consequence of these phases manifest themselves in several predictions. 
The decay with time of the ensemble  energy, Eq. 
(\ref{eqNI01}),
goes through a qualitative change at the boundary between integrability  
and non-integrability $\gamma=1/3$. 
Similarly, for the relation between the mean of the time average and the ensemble average,
Eq. (\ref{eqNI04}).
Finally, the $\mbox{EB}$ parameter, Eq.  
(\ref{eqFLUC10})
and Eq.
(\ref{eqNI13}),
characterises  the fluctuations of the time average
 and it too exhibits a discontinuous behaviour at $\gamma=1/3$. Thus we
have exposed the rich consequences of the fact that an observable
is tuned from being integrable to non-integrable. Interestingly, experiments
use $\alpha=4$ (where energy is non-integrable) and $\alpha=2$ (where energy 
is integrable), so we believe that the classification we performed is of possible practical value. 
Finally, we discovered in Sec. 
\ref{Sec7}
 another sort of transition. For $\gamma<1/4$ the PDF
$P(\Upsilon)$ exhibits  an accumulation effect, blowing up
at $\Upsilon\to 0$, see Fig. \ref{a6}. This implies that some of the particles
remain in the very cold phase, in the sense of very small velocities, 
 for very long periods.

$$ $$
{\bf Acknowledgements}
The support of Israel Science Foundation's grant 1614/21 is acknowledged 
(EB). 
This work was supported by the JSPS KAKENHI Grant No 240 18K03468 (TA).
We thank Tony Albers, Nir Davidson and Lev Khaykovich for helpful suggestions.

\section{Appendix A}

We analyse, laser cooled atoms following the method of Bertin and
 Bardou \cite{Bertin}.
One idea is to analyse the dynamics of the lifetime $\tau(v)$,
taken as a state variable instead of the velocity as done in the main text.
As mentioned in the main text, the process in the time interval 
$(0,t)$ is characterised by a set of uncorrelated speeds 
$(v_1, .... v_N)$ 
all
drawn from the common PDF $f(v)$.
Here $N(t)$ is the random number of
velocity updates (collision events) in the time interval $(0,t)$.
These velocity updates are taking place at random times $(t_0, t_1, ... t_{N-1})$,
and $t_0=0$ is the origin of time. 
The waiting times $\tilde{\tau}_i = t_{i}-t_{i-1}$ are drawn from an exponential
PDF $q(\tilde{\tau}|v)$, Eq. (\ref{eqEXP}),
 defined by the lifetime $\tau(v_i)$. The lifetimes
are thus fluctuating: every update of the velocity implies a
modification of the lifetime.
 We have the sequence of lifetimes
$(\tau_1(v_1), \tau_2(v_2), \cdots)$ and this is a useful
characteristic of the process.  Given the dependence of
the lifetime on $v$, namely given the function $\tau(v)$, then,
if we find the PDF of the lifetime
at time $t$, we can predict the velocity PDF. 

The bare PDF of the lifetimes is given by the chain rule $\psi(\tau)=f(v)|{\rm d} v/{\rm d} \tau|$. More precisely this is the PDF  of the lifetime, immediately after a collision event. It, of course, differs from the PDF of the lifetime
at time $t$, which is denoted 
 $P(\tau,t)$. At time $t$, it is  more likely to find
an atom with a long lifetime compared with a short one (if you arrive
at a bus station randomly, you are more likely to fall on a long time
interval between bus arrivals, if compared to short ones).  
For $\tau(v) \propto  v^{-\alpha}$  for $v \to 0$ the chain rule gives
$$ \psi(\tau) \propto \tau^{-1 - \gamma}, \ \mbox{and} \ \gamma=1/\alpha $$
As mentioned, we assume
 $f(0)\neq 0$,  namely we assume that a particle can be injected at small
speed values. 
If $\alpha>1$ we have a diverging mean lifetime. 
Of course, the PDF of lifetimes $\psi(\tau)$ is not the same as the PDF of the waiting
times $\phi_1(\tilde{\tau})$ discussed in the main text, though both share the same type
of power law decay. 

 The master equation for the lifetime PDF is
\begin{equation}
{\partial P \over \partial t}  = - { P \over \tau} + \psi(\tau) \int_0 ^\infty { P (\tau',t) \over \tau'} {\rm d} \tau'.
\label{eqA00}
\end{equation}
Here both in the loss and the gain terms $1/\tau$ is the rate of leaving
state $\tau$.  In this equation $f(v)$ appears indirectly through $\psi(\tau)$.
In equilibrium, namely $\gamma>1$ we have
\begin{equation}
\lim_{t \to \infty} P(\tau,t) = { \tau \psi(\tau) \over \langle \tau \rangle},
\label{eqA01}
\end{equation}
where the mean is $\langle \tau \rangle = \int_0 ^\infty \tau \psi(\tau) {\rm d} \tau$. 
The fact that we multiply $\psi(\tau)$ with $\tau$ means that in equilibrium we favour the sampling of larger lifetimes, compared to those distributed
with the bare PDF $\psi(\tau)$. When $\gamma<1$ the normalisation
$\langle \tau \rangle$ diverges. Instead we replace the mean with an
effective average $\langle \tau \rangle_{\rm eff} = \int^t \tau \psi(\tau)
{\rm d} \tau \propto t^{1 - \gamma}$. 
Then inspired by Eq. (\ref{eqA01}) we expect
\begin{equation}
P(\tau, t) \sim b_1 { \tau \psi(\tau) \over t^{1-\gamma} },
\label{eqA02}
\end{equation}
where $b_1$ needs a calculation. 
Already from these arguments we expect to find an infinite density
 for the lifetimes
\begin{equation}
\lim_{t\to \infty} t^{1-\gamma} P(\tau,t) = b_1 \tau \psi(\tau)= {\cal I}_\tau(\tau). 
\label{eqA03}
\end{equation}
Here the area under the function ${\cal I}_\tau (\tau)$ clearly diverges
since the mean lifetime is infinite. From here we may find the infinite
density of the velocity, using the chain rule. Namely
$ \rho(v,t) = P(\tau, t ) |{\rm d} \tau(v)/{\rm d} v|$ and then using
Eq. (\ref{eqA03}) we get Eq. 
(\ref{eq12}) (besides a prefactor which we still need to obtain).
Note that
the infinite densities  ${\cal I}_\tau (\tau)$ or  $ {\cal I}_v(v)$  are
 non-normalised due to their large or small argument behaviour respectively.

To solve Eq. (\ref{eqA00}) we introduce  the Laplace transform
\begin{equation}
\hat{P}(\tau,p) = \int_0 ^\infty e^{ - p t} P(\tau, t) {\rm d } t
\label{eqA04}
\end{equation}
where we use the convention that the argument in the parentheses, i.e.
$p$ or $t$, defines  the space we are working in. Using the initial
condition, we have $P(\tau,t)=\psi(\tau)$ at time $t=0$ and
 hence the Laplace transform of 
Eq. (\ref{eqA00}) gives
\begin{equation}
p \hat{P}(\tau,p) - \psi(
\tau) = - { \hat{P}(\tau,p) \over \tau}  + \psi(\tau) 
\int_0 ^\infty {\rm d} \tau' { \hat{P} (\tau',p) \over \tau'} .
\label{eqA05}
\end{equation}
Using the normalisation condition $\int_0 ^\infty \hat{P}(\tau,p) {\rm \tau}= 1/p$ after some straight-forward rearrangement we find
\begin{equation}
K(\tau,p) = \psi(\tau) \int_0 ^\infty K(\tau',p) {\rm d} \tau',
\label{eqA06}
\end{equation}
where $K(\tau,p) = \hat{P}(\tau,p) (p + 1/\tau)$. From Eq. (\ref{eqA06}) we have
$K(\tau,p) = \psi(\tau) h(p)$, where we use the fact that $\psi(\tau)$
is normalised. Hence we get
\begin{equation}
\hat{P}(\tau,p) = {\psi(\tau) \over p + 1/\tau} h(p),
\label{eqA07}
\end{equation}
To determine $h(p)$ we use normalisation $\int_0 ^\infty \hat{P}(\tau,p) {\rm d} \tau = 1/p$. We thus recover the exact  result in \cite{Bertin}
\begin{equation}
\hat{P}(\tau,p) = {1 \over p \tau^{*} (p) } { \tau \psi(\tau) \over 1 + p \tau } \ \
\mbox{with} \ \ \ 
\tau^{*} (p) = \int_0 ^\infty { \tau' \psi(\tau') {\rm d}\tau' \over 1 + p \tau'}. 
\label{eqA08}
\end{equation}

We analyse two cases. The first corresponds to the class of  PDFs of lifetimes
$\psi(\tau)$  with a finite mean and hence describing a stationary process,
and secondly  those PDFs with an infinite mean, namely $0<\gamma<1$. 
The Laplace transform of the waiting time
PDF, for small $p$ gives  \cite{Godreche2001} 
\begin{equation}
\hat{\psi}(p) \sim 
\left\{
\begin{array}{l} 
1 - p \langle \tau \rangle \ \mbox{Case 1} \ \  \mbox{when} \ \langle \tau\rangle \
\mbox{is finite} \\
1 - \tilde{b}_\gamma p^\gamma \ \ \mbox{Case 2} \ \  \mbox{if} \ \  \gamma<1.
\end{array}
\right. 
\label{eqA09}
\end{equation}
The leading term comes from the  normalisation condition.
In the second case, if $\psi(\tau) \sim \gamma A \tau^{- 1 - \gamma}$ with some
amplitude $A$ then $\tilde{b}_\gamma= A \Gamma(1 - \gamma)$, hence as
well known \cite{CT} 
  the far tail
of the waiting time PDF determines the small $p$ behaviour of its corresponding Laplace transform. 
 The amplitude $A$ is related to the PDF of velocities:
using
$\tau(v)\sim c v^{-\alpha}$ for $v \to 0$ 
and the chain rule $\psi(\tau)= f(v)|{\rm d} v / {\rm d} \tau|$,
we get
$A= f(v)|_{v=0} c^{1/\gamma}$ and recall $\gamma=1/\alpha$. 

We wish to investigate $P(\tau,t)$ in the long time limit, hence we
analyse Eq. 
(\ref{eqA08})
in the small $p$ domain. We find 
\begin{equation}
\tau^{*} (p) \sim \left\{
\begin{array}{l l} 
\langle \tau \rangle \ \  & \ \mbox{Case 1} \\
\Gamma(1+\gamma) \tilde{b}_\gamma p^{\gamma-1} \ \ &  \ \mbox{Case 2}.
\end{array}
\right.
\label{eqA10}
\end{equation}
Inserting in Eq. (\ref{eqA08}) we find in the $p\to 0$ limit
\begin{equation}
\hat{P}\left( \tau, p \right) \sim \left\{
\begin{array}{l l}
{ \tau \psi(\tau) \over p \langle \tau \rangle} \ &  \ \  \mbox{Case 1} \\
{ \tau \psi(\tau) \over \gamma \tilde{b}_\gamma  p^\gamma} \ &  \ \ \mbox{Case 2.}
\end{array}
\right. 
\label{eqA11}
\end{equation}
Inverting to the time domain, we find in the long time limit
for both cases
\begin{equation}
P\left( \tau , t\right) \sim {  \tau \psi(\tau) \over \langle \tau(t) \rangle_{{\rm eff}} }, 
\label{eqA12}
\end{equation}
where the effective average waiting time is 
$\langle \tau(t) \rangle_{{\rm eff}}= \langle \tau \rangle$ for case
$1$ and 
$\langle \tau(t) \rangle_{{\rm eff}}= \gamma \Gamma^2( \gamma) \tilde{b}_\gamma t^{1-\gamma}$ for case $2$. In other words we found $b_1$ in Eq. 
(\ref{eqA03}). In terms of the amplitude $A$ we have
\begin{equation}
\lim_{t \to \infty} t^{1 - \gamma} P(\tau,t) = 
{ \sin( \gamma \pi) \over \Gamma(1+ \gamma) \pi} { \tau \psi(\tau) \over A} = {\cal I}_\tau (\tau). 
\label{eqA13}
\end{equation}
This gives the infinite density of $v$ with the chain rules ${\cal I}_{v} (v)
= {\cal I}_\tau (\tau) |{\rm d} \tau/ {\rm d} v|$,  and $\psi(\tau)= f(v)|{\rm d}  v/{\rm d} \tau|$ (recall here
 $A= f(v)|_{v=0} c^{1/\gamma}$ and  $\tau(v)\sim c v^{-1/\gamma}$ for $v \to 0$). 

So far we considered the limit of $p\to 0$ (which is the same
as  $t\to \infty$) while we kept $\tau$ fixed. We saw that, when $\gamma<1$,
we get a non-normalisable solution.  We expect that
at least for this scale free case we can find a second type of scaling solution.
So now we will stick to case
number two only.  We consider a second type of 
long time limit, where in Laplace space the product $p \tau$ remains finite,
while $p\to 0$ and $\tau\to \infty$. Unlike the previous method we will
obtain in this case a normalisable scaling solution. Large lifetimes
correspond to small speeds, and hence to cooling.

 In this limit we need to invert, from the Laplace domain
$s$ to the time domain $t$ the following expression
\begin{equation}
\hat{P}(p,\tau)  \sim {\tau \psi(\tau) \over \Gamma(1 + \gamma) \tilde{b}_\gamma}
{1 \over p^\gamma} {1 \over 1 + p \tau}
\label{eqA14}
\end{equation}
where we used Eqs.   
(\ref{eqA08},\ref{eqA10}). 
Now by definition when $\tau$ is large $[\tau \psi(\tau)]/(\gamma \tilde{b}_\gamma)= 
\tau^{-\gamma} /\Gamma(1-\gamma)$. 
We further use the following triplet of  Laplace pairs
\begin{equation}
{ 1 \over p^\gamma} \leftrightarrow {t^{\gamma-1} \over \Gamma(\gamma)}, \ \ \ \ \ 
{1 \over 1 + p \tau} \leftrightarrow { 1 \over \tau} \exp( - t/\tau),
\ \ \ \ \  {1 \over p^\gamma} {1 \over 1 + p \tau}
\leftrightarrow {1 \over \tau
 \Gamma(\gamma) } 
\int_0 ^t { \exp ( - \tilde{t} /\tau) \over (t - \tilde{t})^{1 - \gamma}}
{\rm d} \tilde{t} , 
\label{eqA14}
\end{equation}
in particular we used the  convolution theorem. 
With a straight forward  change of variables we find the scaling function, namely 
\begin{equation}
P(\tau,t )\sim {1 \over t} \phi\left( {t \over \tau} \right) 
\label{eqA15}
\end{equation}
with
\begin{equation}
\phi(x) = { \sin( \gamma \pi) \over \Gamma(\gamma)  \pi} x^{1 + \gamma} \exp(-x) \int_0 ^1 { \exp( x z) \over z^{1-\gamma} } {\rm d} z.
\label{eqA16}
\end{equation}
The integral can be expressed in terms of incomplete Gamma
functions, and in this sense we have an exact expression
for the scaling function of the random variable $\tau/t$. 
As mentioned, with  this solution we can  predict the scaling behaviour
of the distribution of the speed $v$. Simply change variables
according to $\tau=c/v^{1/\gamma}$ and then use Eq. (\ref{eqA15})
to get Eq. 
(\ref{eq11}). 
Thus as expected  the two methods of solution
yield the same result. 
Note the normalisation reads $\int_0 ^\infty \phi(1/y) {\rm d} y = 1$.

\section{Appendix B} 

\subsection{Basics of renewal theory}

 We start with a brief recapitulation of renewal theory \cite{Godreche2001,Wanli,Cox,Frey,BookR}. Let $\phi_1(\tilde{\tau})$
be the PDF of time intervals between renewal events (in our case collisions
that modify the velocity). The process starts at time $t=0$, we draw a waiting
time from the mentioned PDF, and this defines the point on the time axis for
the first renewal event.  We continue this way for the second event, etc.
In the time interval $(0,t)$ we have $N$ events and the latter is of course
a random variable. Let $Q_N(t) {\rm d} t$ be the probability that the $N$-th renewal
event is taking place in the interval $(t,t+dt)$. Then from the  renewal property
of the process
we have 
\begin{equation}
Q_{N+1} (t) = \int_0 ^t Q_N(t-\tilde{\tau}) \phi_1 (\tilde{\tau}) {\rm d} \tilde{\tau},
\label{eqApB01}
\end{equation}
with the initial condition $Q_0(t)=\delta(t)$.  
The probability of finding $N$ renewals in $(0,t)$ is 
\begin{equation}
P_N(t) = \int_0 ^t Q_N(t - \tilde{\tau} ) W (\tilde{\tau}) {\rm d} \tilde{\tau} 
\label{eqApB02}
\end{equation}
Here $W(t)= 1 - \int_0 ^t \phi_1 (\tilde{\tau})$ is the probability of not making
a transition up to time $t$. 
Eq. (\ref{eqApB02}) thus describes a situation where the $N$th renewal takes place at
time $t-\tilde{\tau}$  and in the remaining
time $\tilde{\tau}$ no jump was made. We now consider the Laplace transform of $P_N(t)$ denoted
$\hat{P}_N(p)$ using Eqs. (\ref{eqApB01},\ref{eqApB02}) and the convolution theorem we
find
\begin{equation}
\hat{P}_N (p) = { 1 - \hat{\phi}_1(p) \over p} [\hat{\phi}_1 (p)]^N.
\label{eqApB03}
\end{equation}
It follows that the mean of $N$ is
\begin{equation}
\langle \hat{N}(p)  \rangle  = { 1 - \hat{\phi}_1(p) \over p}  \sum_{N=0} ^\infty 
N [\hat{\phi}_1 (p)]^N = {1 \over p } { \hat{\phi}_1(p) \over  1- \hat{\phi}_1(p)}. 
\label{eqApB04}
\end{equation}
To analyse the long time limit of $\langle N(t)\rangle$ we investigate $\langle \hat{N}(p) \rangle$ for
small $p$. We are interested in the cases where the mean waiting time diverges, namely following
Eq.
(\ref{ctrw06})
\begin{equation}
\phi_1 ( \tilde{\tau} ) \sim {1 \over \alpha} \tau_{0} ^{1/\alpha} {1 \over \tilde{\tau}^{1+ 1/\alpha} },
\label{eqApB05}
\end{equation}
where in our case $\tau_0 =c [\Gamma(1 +1/\alpha)/v_{{\rm max}}]^\alpha$.
Then in the limit $p \to 0$  one can show that for  $\alpha>1$ 
\begin{equation}
\hat{\phi}_1(p) \sim 1 - |\Gamma(1 - {1 \over \alpha} )| \tau_{0} ^{1/\alpha} p^{1/\alpha} 
\label{eqApB06}
\end{equation}
where the leading term is the normalization. 
Inserting Eq. (\ref{eqApB06}) in Eq. (\ref{eqApB04}) and then performing a
straight forward inverse Laplace
transform one finds $\langle N(t) \rangle \sim [\alpha \sin(\pi/\alpha) / \pi] (t /\tau_0)^{1/\alpha}$ \cite{Godreche2001}. 
This in turn gives the expression for $\langle N(t) \rangle$ in Eqs. 
(\ref{eq1a3},
\ref{eq1a5}) of the main text. 
The distribution of $N$ in the long time limit is obtained using the
small $p$ behaviour of 
Eq. (\ref{eqApB03}), let us denote $\hat{\phi}_1(p) \sim 1 - b_\gamma p^\gamma$
and then
\begin{equation}
\hat{P}_N(p) = {1 - \hat{\phi}_1 (p) \over p } \exp\left[ N \ln \hat{\phi}_1 (p)\right] \sim b_\gamma p^{\gamma-1} \exp( - N b_\gamma p^\gamma).
\label{eqApB07}
\end{equation}
Inversion is made possible with the same tricks used to derive Eq. 
(\ref{eqFLUC09}) 
\begin{equation}
P_N(t) 
\sim 
{ t \over \gamma b_{\gamma} ^{1/\gamma} N^{1 + 1 /\gamma}} l_{\gamma,1} \left( { t \over ( N b_\gamma)^{1/\gamma}  } \right)
\label{eqApB08}
\end{equation}
where $l_{\gamma,1}(.)$ is the one sided L\'evy PDF. 
Thus the PDF of the action ${\cal S}(t)$ discussed in the main text is the same
as the PDF of number of renewals
$N$ besides a scale and  provided that $1/3<\gamma<1$.

\subsection{Derivation of Eq. (\ref{eq1a3}) }

 We now  explain how to obtain Eq. (\ref{eq1a3}) using Eq. 
(\ref{eqG09}).  We investigate the latter in the limit
$p \to 0$ corresponding to long times. First one can show that the second term on the right hand side
of Eq. (\ref{eqG09}) is negligible provided that $\alpha<3$.  Secondly from the definition 
of the Laplace transform, we have
\begin{equation}
- { \partial \hat{\phi}_2 (u,p) \over \partial u} |_{u=p=0} = \langle s \rangle 
\label{eq1a4}
\end{equation}
and using convolution $\hat{\Phi}_2(0,p) = [ 1 - \hat{\phi}_2(0,p)]/p$. Hence from Eq. (\ref{eqG09}) we
have
\begin{equation}
\langle \hat{S}(p) \rangle \sim { \langle s \rangle \over p \left[ 1 - \hat{\phi}_2(0,p)\right]}. 
\label{eq1a4bb}
\end{equation}
When  $u=0$ the Laplace transform
of the joint PDF $\hat{\phi}_2 (u,p)$  reduced to the Laplace transform of the marginal PDF of
waiting times, namely 
$\hat{\phi}_2 ( 0, p)= \hat{\phi}_1 (p)$. We now use Eq. 
(\ref{eqApB04}) and  $\hat{\phi}_1(p)\sim 1$ to leading order in $p$ 
finding
$\langle \hat{S}(p) \rangle \sim  \langle s \rangle \langle \hat {N} (p) \rangle$,
which gives  Eq. (\ref{eq1a3}).

\end{widetext}

\end{document}